\documentclass{vldb1}

\pdfoutput=1

\usepackage{float,latexsym,enumerate}
\usepackage{hyperref}
\usepackage{graphicx}
\usepackage{tabularx}
\usepackage{alltt}
\usepackage[labelfont=bf]{caption}
\usepackage{xcolor}
\usepackage{balance}
\usepackage{microtype}
\usepackage{amssymb}
\usepackage{amsmath}
\usepackage{algorithm,algorithmic}
\usepackage{mathtools}
\usepackage{subcaption}
\usepackage{enumitem}
\newfloat{algorithm}{t}{lop}

\usepackage{etoolbox}
\makeatletter
\preto{\@verbatim}{\topsep=0pt \partopsep=0pt }
\makeatother

\hypersetup{colorlinks, citecolor=black, filecolor=black, linkcolor=black,urlcolor=blue}
\floatstyle{boxed}
\newfloat{program}{thp}{lop}
\floatname{program}{Program}

\newif\iffullversion
\fullversiontrue

\newcommand{\revision}[1]{{{#1}}}

\iffullversion
\newcommand{\fullversion}[1]{#1}
\newcommand{\paper}[1]{}
\else
\newcommand{\fullversion}[1]{}
\newcommand{\paper}[1]{#1}
\fi

\newcommand{\eat}[1]{}
\newcommand{\smalltt}[1]{{\small{\texttt{#1}}}}
\newcommand{\paras}[1]{\noindent\textit{#1}:}

\author{%
	Bikash Chandra \\
	\small{\em IIT Bombay} \\
	\small{\em bikash@cse.iitb.ac.in}
	\and
	 S. Sudarshan \\
	\small{\em IIT Bombay} \\
	\small{\em sudarsha@cse.iitb.ac.in}
}

\title{Runtime Optimization of Join Location in Parallel Data Management Systems
}

\begin{document}

\maketitle

\begin{abstract}
Applications running on parallel systems often need to join a streaming relation or a 
stored relation with data indexed in a parallel data storage system. Some applications 
also compute UDFs on the joined tuples. The join can be done 
at the data storage nodes, corresponding to reduce side joins, or by 
fetching data from the storage system to compute nodes, corresponding to map side join. 
Both may be suboptimal: reduce side joins may cause skew, while map side joins may lead 
to a lot of data being transferred and replicated. 

In this paper, we present techniques to make runtime decisions between the two options 
on a per key basis, in order to improve the throughput of the join, accounting for UDF 
computation if any.
Our techniques are based on an extended ski-rental algorithm and provide worst-case 
performance guarantees with respect to the optimal point in the space considered by us.  
Our techniques use load balancing taking into 
account the CPU, network and I/O costs as well as the load on compute and storage nodes.
We have implemented our techniques on Hadoop, Spark and the Muppet stream processing 
engine. 
Our experiments show that our optimization techniques provide a significant improvement 
in throughput over existing techniques.
\end{abstract}

\section{Introduction}
\label{sec:intro}
Parallel batch data  processing systems such as Hadoop MapReduce and 
Spark~\cite{spark} 
are designed to process massive amounts of data on a large cluster of machines. Parallel 
stream processing systems, such as Storm~\cite{storm} and Muppet~\cite{muppet}, on the 
other hand, are aimed at processing a fire-hose of data that arrive 
at very fast rates.  These frameworks work on many nodes and are designed to 
serve different ends of the spectrum in terms of latency and throughput. Systems such as 
HBase and Cassandra are often used as backend data stores and support indexed access on 
primary keys.

In this paper, we consider a class of applications that run on such parallel systems and 
need to perform a join of input data, which may be streaming or stored data, with other 
data stored and indexed in a parallel data store.  
The application may also compute a UDF based on the joined tuples. 
There are numerous such applications that compute UDFs on 
join results, such as entity annotation using large stored models on streaming/stored 
textual data and genome sequence read alignment. We discuss a few applications later in 
Section~\ref{sec:motivation}. The focus of this work is on 
joins where the data stored in the parallel data store is indexed on the join 
attributes. (In case the data is not already indexed, an index can be created before 
computing the join.)

\begin{figure}
	\begin{center}
		\includegraphics[width=0.285\textwidth,keepaspectratio=true]{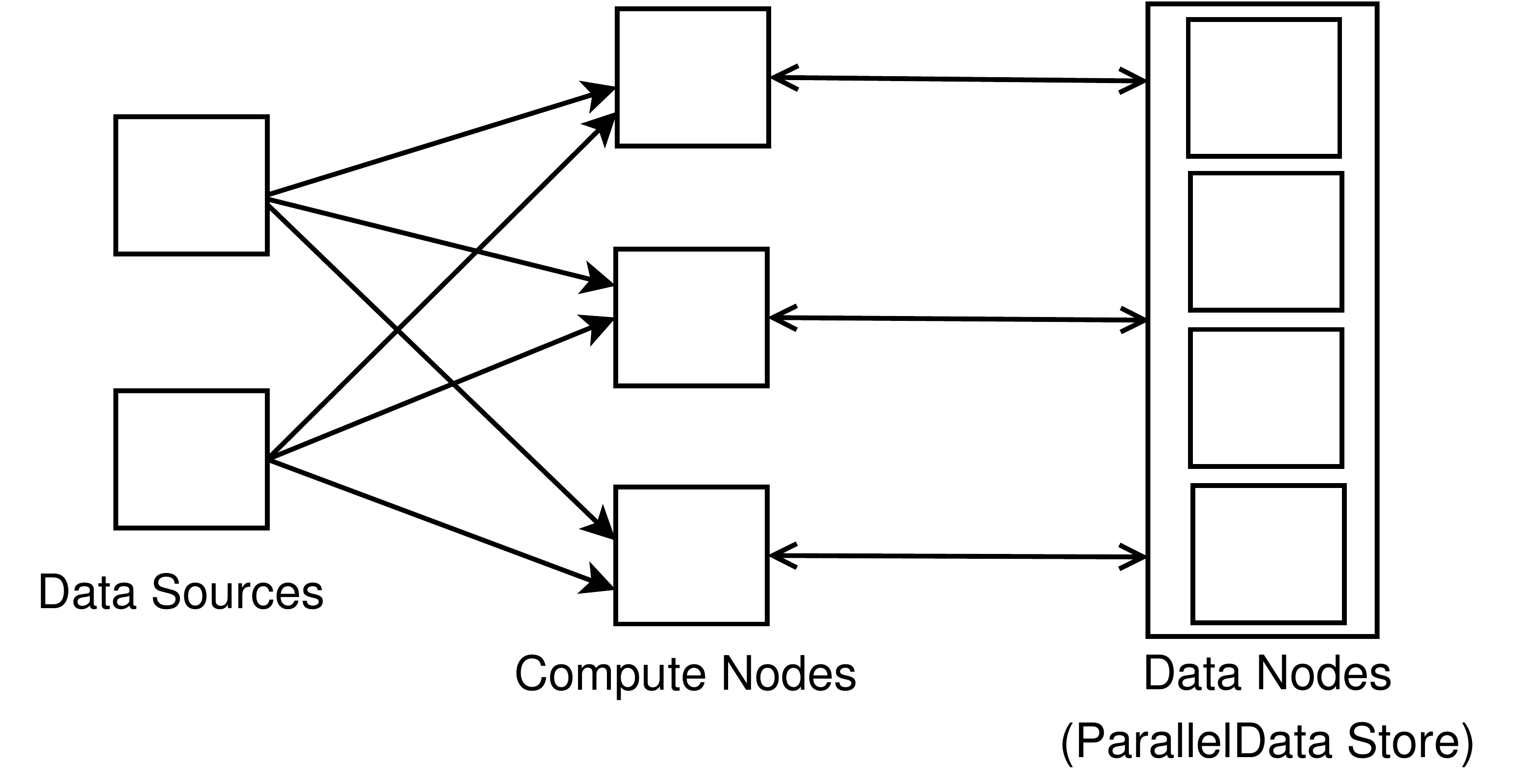}
		\caption{Parallel data management system with datastore access (data sources can 
		have stored or streaming data)}
		\label{fig:goal}
	\end{center}
\end{figure}

We refer to the nodes on which the application is running as 
\textit{compute nodes}, and the nodes on which the data is stored as \textit{data nodes}. 
\eat{Compute nodes and data nodes may be physically co-located on the same machines if 
desired. }
A schematic representation of our architecture is shown in 
Figure~\ref{fig:goal}. 

Joins may be performed in different ways.
\begin{enumerate}[leftmargin=*]
\setlength{\itemsep}{0pt}
\setlength{\parskip}{0.1pt}
 \setlength{\parsep}{0pt}
\item 
One way to perform the join is to fetch data from the data store, for 
each incoming stream/stored relation tuple, and execute the join (and UDF computation, if 
any) at the compute node. 
This is analogous to a map-side join or a parallel index nested-loops join.

With a naive implementation for accessing data from the parallel data store, for each 
request sent to the data store, the compute node must wait for the result to come back, 
and then complete the computation, before moving on to the next data item.
Such an implementation could be very inefficient due to high latencies for fetching data 
from the data store. 
Batching and asynchronous submission can be used to improve performance. However, this 
approach requires more data transfer if the UDF result is smaller than the input. 

\item
An alternative technique is to send the incoming tuples from the compute nodes to the 
data nodes and perform the join (along with the UDF computation, if any) at the data 
node. This technique is analogous to reduce-side join.
Note that parallel hash joins used in parallel relational databases also correspond to 
reduce-side join. The build relation can be partitioned and indexed at the data nodes 
while nodes with the probe relation can be used as compute nodes.

A drawback of the reduce side join approach is that it is vulnerable to skew
if the input tuples have keys that are very frequent (heavy 
hitters) or the cost of UDF computation is higher for some key values. 
Such skew could result in some data nodes performing significantly more computations 
compared to other data nodes.
Existing techniques to handle skew in parallel joins are based on using statistics or 
sampling to find balanced partitioning boundaries;  in addition, some techniques such as 
\cite{skew:seshadri,flowjoin}
also use statistics to find and replicate heavy hitters (very frequent keys). As 
described later in Section~\ref{sec:entity}, 
these techniques can be extended to mitigate skew in entity annotation application. 
However, the techniques proposed earlier require a fixed threshold parameter to determine 
which keys are heavy hitters. In a streaming setting, statistics may be unavailable prior 
to execution and may change over time. Another technique to mitigate skew is to 
dynamically reallocate partitions to less loaded machines. However, this may be 
ineffective against skew caused by a single key or few keys. 
\end{enumerate}

A key idea in this paper is to process incoming tuples by dynamically deciding whether to 
perform the join by fetching values from the parallel data store (corresponding to 
map-side join), or by sending values to the parallel data store (corresponding to 
reduce-side join). One of these alternatives is chosen at runtime on a per join-key 
basis, depending on the frequency of the key and on the load at the compute and data 
nodes. Our goal is to optimize data access and computation, and minimize skew in this 
setting, in order to improve throughput.\footnote{Throughput is the number of input 
tuples processed per unit time.}

Our main contributions in this paper are as follows.
\begin{enumerate}[leftmargin=*] 
\setlength{\itemsep}{0pt}
\setlength{\parskip}{0.1pt}
 \setlength{\parsep}{0pt}
 \item We address the issue of dynamically deciding whether to perform the join at 
 the compute or the data node using a ski-rental formulation. We give an overview of our 
 framework in Section~\ref{sec:framework}. We then present, in Section~\ref{sec:ski}, 
 generalizations of the classical ski-rental~\cite{ski-rental} problem along with 
 worst-case performance guarantees. The generalized ski-rental is used to decide which 
 items from the data store are being accessed frequently enough to be fetched from the 
 data store and cached locally.

 \item In Section~\ref{sec:load_balance}, we present techniques that allow the 
 computation to be balanced between compute nodes and data nodes with the goal of 
 maximizing the throughput.
 
 \item Our techniques are able to mitigate skew without any pre-computed statistics 
 and work well with batch systems as well as with streaming systems, where statistics may 
 not be available, and heavy hitters may change over time. Using our technique, heavy 
 hitter keys  would end up being cached across all compute nodes and would be processed 
 at the compute  nodes. Requests for infrequent keys, on the other hand, are sent to the 
 data node and can be processed at the data node. 
 Our techniques thus represent an alternative to existing techniques 
 to handle hash join skew based on statistics, such as \cite{skew:seshadri,flowjoin}. 
 
 Our techniques allow dynamic addition and deletion of compute nodes since the 
 compute nodes do not store intermediate join results or any state (other than cached 
 data). In the context of streaming data, such elasticity can be used to add resources 
 to handle peak load, while using less resources at low load.

 \item Our runtime techniques can be used in the context of multiple joins (as described 
  in  Section~\ref{sec:multiple}), where statistics of intermediate join results may not 
  be  available.

 \item  We have also extended the APIs of the MapReduce, Spark and Muppet frameworks to 
 incorporate the  techniques of batching and prefetching. We describe these extensions 
 in Section~\ref{sec:other_opt}. Our techniques can also be used in traditional parallel 
 relational databases.

\item We conduct several experiments, as described in Section~\ref{sec:expt}, based on an 
entity annotation application as well as using some synthetic applications. We show that 
our optimization techniques provide significantly better throughput compared to 
alternatives for both batch processing and stream processing. In fact, our techniques 
perform better than a custom partitioner described in~\cite{soumen:hipc} to mitigate skew 
for the entity annotation implementation.
\end{enumerate}

We provide a detailed comparison with related work in Section~\ref{sec:relwork}.

\section{Motivating Applications}
\label{sec:motivation}

In this section, we use the example of entity annotation to motivate the need for our 
framework and also as a running example throughout 
this paper to demonstrate our optimization techniques. We also discuss other motivating 
applications.

\subsection{Entity Annotation}
\label{sec:entity}

Entity annotation is the problem of marking up words in a document with the entities that 
they refer to. Entity annotation can be done using trained machine learning models.
For each ``token'', i.e. name/word, a model is precomputed and stored, indexed by the 
token.
To annotate a document, in the first step, tokens which are possible mentions of entities 
are identified, e.g. `Michael Jordan', along with the surrounding text. 
In the next step, a classification function is applied on the model corresponding to the 
token along with text surrounding the token, to identify the entity which the token 
refers to: `Michael Jordan', the basketball player or  `Michael Jordan', the computer 
scientist and professor.

A reduce-side join can be used for entity annotation. 
Models are partitioned amongst reducers. 
A mapper extracts tokens, with surrounding text, and maps it based on 
tokens. The reducer joins the token with the models and performs classification using 
models. 

\begin{figure}
\begin{scriptsize}
\begin{verbatim}
map(docId,document) {
 for each spot in document.getSpots(){
  spotContext = getContextRecord(spot,document)
  context.write(spotContext.key, spotContext.value)
 }
}
partitioner(key) {
 if(isFrequent(key)) {
  //spread across multiple reducers
  return randomPartition
 } else {
  //to localise access to stored models
  return getPartition(key)
 }
}
reduce(key,values) {
 model = getModel(key)
 for each spotContext in values {
  annotatedValues = 
    classifyRecord(spotContext, model)
  context.write(annotedValues)
 }
}
\end{verbatim}
\end{scriptsize} 
\caption{Entity annotations with join at reduce}
\label{fig:mapReduceEntity}
\end{figure}
\begin{figure}
\begin{scriptsize}
\begin{verbatim}
map(docId,document) {
 for each spot in document.getSpots(){
  spotContext=getContextRecord(spot,document)
  annotatedValues = 
      f(spotContext.key,spotContext.value)
  context.write(spotContext.key, annotatedValues)
 }
}
f(key,params) {
 model = modelStore.getModel(key)
 annotatedValues = classifyRecord(params,model)
 return annotatedValues
}
\end{verbatim}
\end{scriptsize}
\caption{Entity annotations with map-side joins}
\label{fig:mapEntity}
\end{figure}

Gupta et al.\ show in \cite{soumen:hipc} that entity annotation using reduce-side join 
is inefficient because of skew, and present a custom partitioning function to mitigate 
skew due to both key frequency and classification function computation cost imbalance 
across different models.
To reduce skew, models with high costs due to frequent tokens or high classification 
cost are replicated across all partitions; models with low costs are kept at one 
partition. Pseudo code for their approach is shown in Figure~\ref{fig:mapReduceEntity}. 
In the map function, \smalltt{spotContext.key} is the token and  
\smalltt{spotContext.value} is the surrounding text. Frequent tokens are routed randomly 
to any one of the reducers while non frequent ones are routed to the same partition as 
the model for the token. Their solution may be viewed as an extension of the work by 
DeWitt et al.\ \cite{skew:seshadri}, for minimizing parallel join skew, since they use 
a similar partitioning/replication technique, but \cite{soumen:hipc} also take the 
computation time for the classification function into account.

Entity annotation can also be done at the map side by fetching models to the mapper 
and performing the annotations at the mapper. Sample pseudocode for this approach is 
shown in Figure~\ref{fig:mapEntity}, where $f(key,params)$ is the function that performs 
the entity annotation. However, if done naively, this approach 
will lead to a lot of data being transferred from the data store and data accesses would 
be blocking, making it inefficient. 

Our approach performs the join as part of the map function but, for each token, 
chooses between fetching the model and performing annotation at the mapper, versus 
sending the token with surrounding text to the nodes hosting the models. 

Our approach can also be used to annotate entities in a text stream. For example, entity 
annotation is important for Twitter streams~\cite{annotate:twitter}. For a Twitter 
stream, new events which did not exist earlier may suddenly gain popularity. Hence, it 
may not be possible to use pre-computed statistics to decide frequent keys. On the 
other hand, our approach does not assume any distribution, but computes statistics at 
runtime, allowing it to adapt to changes.

\subsection{Other Applications}
In general, our framework can be applied to applications that perform joins 
in a batch setting. It can also be used for stream-relation joins in a parallel setting. 

Many machine learning models can make use of the parameter server framework described in
\cite{smola:parameter} for efficient distributed implementation. In this framework, 
machine learning models can be represented as a set of $<$key,value$>$ pairs and can be 
shared across multiple nodes for parallel execution. Li et al.\ \cite{li:parameter} show 
how batched access (which they define as range push and pull) and asynchronous tasks can 
lead to more efficient learning algorithms. 
Our techniques perform ski-rental based caching and dynamic load balancing in addition to 
batching and asynchronous  calls.

CloudBurst~\cite{cloudburst} aligns a set of genome sequence reads with a reference 
sequence using MapReduce, executing a UDF as part of the reduce operation.
Our framework could be used in this case to mitigate skew 
and reduce the runtime; details can be found in \paper{our full paper 
\cite{throughputOpt:arxiv}.} 
\fullversion{Appendix~\ref{app:applications}.}
Our techniques can also be used to minimize skew for joins parallel database settings. 
\section{Framework and Solution \\Approach}
\label{sec:framework}

In this section, we give an overview of our general framework. We also mention how cost 
measurements required for our optimizations are computed.

\subsection{Framework}

We now consider a general framework for applications whose throughput can be optimized 
using our techniques. The application runs on a parallel data management system 
and needs to process a join of a stored relation or stream with stored data, and 
optionally perform some computations based on the join. 
Our approach is to place the stored data, indexed on the join key, on a parallel data 
store. To process the incoming data item, we look up values from the parallel data store 
using the join key and perform computations if any based on the fetched value. 
If the remote data is not already indexed by the join key on a parallel data store, it 
can be repartitioned, an index can be built and our techniques can then be used.

Many parallel data stores support execution of user-defined functions for specific data 
items at the data nodes e.g. endpoints in HBase. The ability to execute user-defined 
functions at the data nodes enables us to push function execution to the data 
nodes. In this paper, we consider only side-effect free functions, which
allows us to perform the function at the data nodes or compute nodes.\footnote{
Extensions to handle special cases of functions with side effects are possible and are an 
area of future work.} 

The function can be considered to be of the form $f (k, p)$, where 
$k$ is a key, obtained from the incoming data item, which is used to fetch valued from 
the data store, and $p$ is a list of other parameters to the function.
\revision{The function $f(k,p)$ first 
fetches the value $v$ stored in the data store for some relation $r$, corresponding to 
the key $k$. The function $f$ then invokes a UDF $f'(k,p,v)$ which can be used to perform 
the desired computations.} The list of parameters, $p$ can be empty and the function can 
merely return the stored value in case no computation is to be performed.

The functions $f(k,p)$ can be invoked in one of several ways.
\begin{enumerate}[leftmargin=*]
\setlength{\itemsep}{0pt}
\setlength{\parskip}{0.1pt}
 \setlength{\parsep}{0pt}
\item For each key $k_i$, fetch the stored value $v_i$, and compute $f'(k_i,p_i,v_i)$ at 
the compute node; $v_i$ can be cached for future computations with different $p$ values. 
The request to fetch the stored data node is called a \textit{data request}. 
\item Send values ($k_i,p_i$) to the data node with value $k_i$, and 
compute  $f'(k_i,p_i,v_i)$ at the data node. This corresponds to stored procedure 
invocations in a database  or coprocessor invocations in HBase. The request for invoking 
functions on a data node is called a \textit{compute request}.
\item Decide on 1 or 2 dynamically, based on parameters such as the sizes of $p$, $v$, 
the cost of computing $f$, the number of invocations on a key $k_i$ at each compute 
node, and the load on the data node storing $k_i$.
\item Send each request individually, or in batches. Sending each request synchronously 
may lead to blocking waits and hence we may issue prefetch requests which could also be 
batched. There are multiple ways in which batched prefetching can be done. We discuss our 
approach in Section~\ref{sec:other_opt}. 
\end{enumerate}

The optimization goal, when choosing from the above alternatives, is to maximize the 
throughput of the system, i.e. the number of $f(k,p)$ invocations that are handled in a 
given amount of time; when the entire set of values $(k,p)$ are already available (i.e. 
in a batch setting), the above goal minimizes the total time to completion, while in a 
streaming setting, the goal directly maximizes throughput. 
\revision{We note that our main optimization techniques are only applicable when user 
defined functions can be executed on data nodes. Most popular data stores like HBase, 
Cassandra and Amazon Redshift do support execution of user defined functions.}

In this paper, we look at online optimization, i.e., the optimization decision is made 
during runtime without any precomputed statistics and without any knowledge of the future 
computations at any given point in time. Instead, we compute statistics at runtime and 
make decisions as follows.
\begin{itemize}[leftmargin=*] 
\setlength{\itemsep}{0pt}
\setlength{\parskip}{0.1pt}
 \setlength{\parsep}{0pt}	
\item Decisions on data requests vs.\ compute requests, and on caching,
are made based on the observed frequency of access.  Our techniques for
making these decisions are described in Section~\ref{sec:ski}.
\item Decisions on load balancing between compute and data nodes are made 
based on the observed load at the compute nodes and data nodes, taking
network bandwidth also into account.  Our techniques for making these
decisions are described in Section~\ref{sec:load_balance}.
\end{itemize}

We assume that the underlying application or streaming data system ensures that
the compute load is balanced across the compute nodes; for example, the input
data could be distributed in a round-robin fashion to ensure load is balanced.
Thus, skew due to data distribution at compute nodes is likely to be small. 

We also assume that the stored data is distributed across data nodes in such a way
that long term load is balanced.  Data storage systems can perform data migration
to deal with load imbalances across data nodes, but since data migration is
usually expensive, this would be done for long-term load imbalances. 
Our caching techniques help minimize the skew due to repeated requests for 
the same key values, which is particularly useful with heavy-hitter skew.
Our techniques for load balancing of computation between data and compute nodes
can reduce skew at data nodes by transferring part of the UDF computation load to
compute nodes.

\subsection{Parameters for Cost Computation}
\label{sec:cost}

In order to decide between compute and data requests, we need to take into account
the CPU, disk and network costs for data access and function execution. We neglect the 
cost of memory access since it is small as compared to disk and network costs. The 
parameters that we take into account for cost computations are listed in 
Table~\ref{tab:cost}. We normalize all costs to the unit of time. \revision{Instead, we 
measure cost parameters at runtime.}

Our implementation measures disk and CPU costs at runtime.  Measurement of network 
bandwidth is done prior to execution; details are provided in 
\paper{\cite{throughputOpt:arxiv}.} 
\fullversion{Appendix~\ref{app:bandwidth}.}
Disk, CPU and network costs may change over time.  To accommodate these changes, these estimates are initialized once and then 
updated periodically based on the actual values. We also need to guard against temporary 
spikes (for e.g. due to changes in system load) in these values. Hence, we perform 
exponential smoothening using the formula
\begin{equation*}
value_{t+1}=\alpha\, \, value_{measured}+(1-\alpha)value_{t}
\end{equation*}
where $\alpha$ is a parameter between 0 and 1. 

\begin{table} 
\begin{center}
\caption{Parameters for cost computation}
  \begin{tabular}{|l|p{6.4cm}|} \hline
    $netBw_i$ & effective network bandwidth $i$\\ \hline
   $s_v$ & size of stored item $v$\\ \hline
     $s_p$ &  average size of parameters \\ \hline
    $s_k$ &  average size of key \\ \hline
   $s_{cv}$ & average size of computed values  \\ \hline
   $tDisk_i$  & average time taken to fetch record from disk at node $i$ \\ \hline
  $tc_i$ & average CPU time taken to compute the function at node $i$ \\ \hline
\end{tabular}
\label{tab:cost}
\end{center}
\end{table}

\section{Frequency Based Runtime Optimization}
\label{sec:ski}
In this section, we look at runtime optimization based on the frequency of access of 
values from the data store to decide between data requests and compute requests. For 
the function $f(k,p)$, the number of calls for each $k$ may be 
different. In typical usage scenarios like web click log 
analysis and annotating entities, the number of calls for each $k$ tend to be very 
skewed.

Once a data item corresponding to a key k is fetched, it can be cached and used with 
different values of $p$. Our focus, initially in this section, is on deciding 
between data and compute requests. Caching of values fetched from the data nodes is  
discussed later in this section. 

\subsection{Basic Ski-Rental}
For the optimization decision of choosing between data requests and compute requests, we 
model the problem as a ski-rental problem. The classical ski-rental 
problem~\cite{ski-rental}  refers to 
the online problem in which there is a choice between paying a renting cost for each 
usage of the object versus paying once for the purchase of the object after which no cost 
for renting needs to be paid. This is an online problem since the number of times that 
the object will be used is not known beforehand.

Suppose that the cost to rent is $r$ and the cost to purchase is $b$. The basic 
ski-rental solution is to keep renting for the first $b/r$ times and  then purchase the 
object. The cost is never more than double the optimal cost (the cost of an offline 
algorithm) and the competitive ratio is 2.
In our problem setting, compute requests can be considered as renting and fetching the 
values locally can be considered as buying.

\subsection{Extensions to Ski-Rental}
Our problem is different from the classical ski-rental in some key ways. In the classical 
ski-rental formulation, once an item is bought there is no recurring 
cost on using the item. In our problem setting, a recurring CPU cost is incurred even 
after the data corresponding to a key has been fetched. 
Ski rental also does not take into account limited storage for the items bought. In our 
case, we only have limited cache size to store the fetched items. Hence, we may need to 
evict some items that have already been bought. The amount of cache required for each 
item may be different and should also be taken into account when buying.
The classical ski-rental problem also does not take into account updates to items.
We now discuss how to handle these extensions.

\subsubsection{Recurring Costs After Buying}

Let the recurring cost after buying be $b_r$. We should keep renting as long as the 
cumulative renting cost is less than the cumulative buying cost. Let $m$ be the number of 
accesses at which we buy. Then
\begin{equation*}
r*m \leq b + b_r * m
\implies m \leq \frac{b}{r-b_r}   \text{ , if } r>b_r
\end{equation*}
If $r\le b_r$, it is cheaper to always rent.

Thus the item must be bought when the number of accesses for the item is ${b}/({r-b_r})$. 
We denote this number by $M$. 
In the worst case scenario, the item is bought and is no longer accessed in the future. 
In that case, the total cost would have been $r*M+b$ while the optimal cost would have 
been $r*M$. 
Thus, 
\begin{align*}
\text{Competitive ratio } &= \frac{\text{Total cost}}{\text{optimal cost}}= 
(r*M+b)/(r*M)\\
&= 2-\frac{b_r}{r} \text{ , substituting } M = \frac{b}{r-b_r} 
\end{align*}

For $b_r=0$, the formulation reduces to the basic ski-rental and the competitive ratio is 
2.

\subsubsection{Caching}
\label{sec:ski:caching}
We consider two types of caches, a memory cache and a disk cache, to store the items that 
have been bought. Access to memory cache is fast but memory is limited. Hence, not all 
purchased items can be cached in memory.

We denote the recurring cost (fetch from cache and compute) where data is cached in 
memory as $b_{rM}$ while the recurring cost where data is cached on disk is $b_{rD}$. 
Since disk access imposes additional overhead, we assume that $b_{rD} \geq b_{rM}$. We 
denote the memory cache as \smalltt{mCache} and disk cache as \smalltt{dCache}. 

Our caching strategy is described in algorithm \smalltt{skiRentalCa\-ching} which is 
shown in Algorithm~\ref{algo:ski}. The algorithm uses the functions 
\smalltt{updateBe\-nefit} and \smalltt{updateCounter} to update the caching benefits and 
access count for a given key. The access count for each key is used to make ski-rental 
based decisions for compute or data requests, as shown in lines 11 and 14 of the 
algorithm. Caching benefits for each key is used to make caching decisions for the 
\smalltt{condCacheInMemory} function.

\begin{algorithm}
\renewcommand{\algorithmicrequire}{\textbf{Inputs:}}
    \renewcommand{\algorithmicensure}{\textbf{Output:}}
\begin{algorithmic}[1]
\REQUIRE k = data item key

\STATE updateBenefit(k)
\STATE updateCounter(k)
\IF{ k $\in$ mCache}
  \STATE v $\leftarrow$ mCache.get(k)
  \STATE localComputeQueue.add(f,k,p,v)
\ELSIF{ k $\in$ dCache}
	 \STATE v $\leftarrow$ dCache.get(k)
	  \STATE localComputeQueue.add(f,k,p,v)
	  \STATE condCacheInMemory(k,v,itemSize)
\ELSE
	\IF{counter(k)$\leq$ $b/(r-b_{rM})$} 
		\STATE computeQueue.add(f,k,v,p)    
	 \ELSE
	      
	     \IF{condCacheInMemory(k,$\phi$,itemSize)}
	     	 \STATE dataQueue.add(mCache,f,k,p)
	      \ELSIF{{ counter(k)$\leq$ $b/(r-b_{rD})$} }
	     	 \STATE computeQueue.add(f,k,v,p)
	      \ELSE
	     	 
	     	 \STATE dataQueue.add(dCache,f,k,p)
	      \ENDIF	
	 \ENDIF 
\ENDIF
\end{algorithmic}
\caption{ : skiRentalCaching}
\label{algo:ski}
\end{algorithm}

Algorithm~\ref{algo:ski} first checks to see if the requested data item is present in 
\smalltt{mCache} and then \smalltt{dCache}. If the requested data item is in cache, it is 
fetched from cache and added to \smalltt{localComputeQueue}; the function 
\smalltt{condCacheInMemory} (described shortly), decides if the item fetched from 
\smalltt{dCache} is to be cached in memory (Lines 3-9).

In case of a cache miss (i.e. the data item is not found in \smalltt{mCache} or 
\smalltt{dCache}) the algorithm checks to see if the data item should be fetched,  
based on ski-rental, taking into account the recurring cost as $b_{rM}$. 
If the ski-rental condition is not satisfied, the data item is added to 
\smalltt{computeQueue} (Lines 11-12).\footnote{ 
If the ski-rental determines that number of access is not sufficient for memory 
caching, it is also not sufficient for disk caching since  $b_{rD} \geq b_{rM}$.} 
\revision{If the ski-rental condition is satisfied, the algorithm needs to check if there 
is sufficient free space in \smalltt{mCache} to cache the item or if existing items with 
low caching benefit can be evicted to \smalltt{dCache} to make space for the current data 
item. This is done using the \smalltt{condCacheInMemory} (Line 14) described below.}
If the item can be cached in memory, the item is added to \smalltt{dataQueue} (Line 15), 
otherwise the algorithm checks if the disk cache ski-rental condition is satisfied 
based on recurring cost as $b_{rD}$ (Line 16). If the condition is not 
satisfied  the item is added to \smalltt{computeQueue} (Line 17) else it is added to 
\smalltt{dataQueue} (Line 19). 

There are multiple parallel threads that do the following (a) read data items from 
\smalltt{localComputeQueue} to perform the function computations, (b) read data items 
from \smalltt{dataQueue}, fetch values from the data store, cache the data item on the 
appropriate cache and compute the function, and (c) read items from 
\smalltt{computeQueue} and issue compute requests to the data nodes.

The function \smalltt{condCacheInMemory} checks to see if the item (given its benefit) 
should be cached in memory or not, either using the available free space in 
\smalltt{mCache} or by evicting items with less benefit from \smalltt{mCache} to 
\smalltt{dCache}. In case  the second argument to the function call is not $\phi$, it 
also performs memory caching if the decision is positive. 
The decision of whether to evict an item from \smalltt{mCache} to \smalltt{dCache} to 
free up space in \smalltt{mCache} can be made using based on the frequency and recency of 
access using existing cache eviction techniques. There are 
several techniques to implement frequency based cache replacement with aging, as surveyed 
in \cite{cache:survey}. We use the weighted LFU-DA algorithm~\cite{lfuda} which assigns 
benefits to data items in such a way that recent and frequent accesses are assigned more 
benefit. Details of how we implement the \smalltt{condCacheInMemory} function are 
described in \paper{\cite{throughputOpt:arxiv}.} 
\fullversion{Appendix~\ref{app:caching}.}

\subsubsection{Updates to the Data Store}
While the data processing framework is running, there could be updates to the data store. 
Updates to data store are another extension to ski-rental where cached (bought) items 
can no longer be used once they are updated. If an item had been bought and 
then it got updated, the purchased item  must be discarded.  If the item was being 
rented when it was updated, the item should be treated as a new item and the count of its 
accesses should be 
set to 0 to ensure frequently updated items are not bought. 
Note that the worst case 
guarantee (cost is $2-b_r/r$ of the optimal cost) still holds even without setting the 
count to 0, but we would unnecessarily buy items that are frequently updated.

There are two ways in which the cache update and count reset can be done. The data node 
could send a broadcast notification to all compute nodes with the key being updated. 
However, frequent updates may flood the nodes of the system leading to poor performance.
The alternative is for each data node to maintain a record of the compute nodes where 
each of its data item has been fetched and cached. In case of updates, the data node 
sends notifications only to the compute nodes where the item has been cached. 
Relevant compute nodes then invalidate the cache for the particular item. 
This approach does not flood all nodes but may result in nodes which have not yet cached 
an item missing the update notification, thereby not resetting the count.
Therefore, with each response to a compute request, the data node also sends 
the timestamp when the item was last updated. The compute node tracks the timestamp of 
the last compute request for each data item. If the timestamp gets updated 
between two compute requests, the counter for the data item is reset.

HBase coprocessors can be used for providing notifications when a row is updated; these 
notifications can be used to invalidate caches for the corresponding keys. Implementation 
of update notifications is an area of future work.

\subsection{Using Modified Ski-rental}
We now formally put together the ideas discussed earlier in this section.
In our problem setting 
the rental cost,  $tCompute$ corresponds to  the cost of sending the compute request to 
the data node, fetching the value of the stored data locally at the data node, computing 
the function and sending back the computed result to the compute node. 
The purchase cost, $tFetch$ corresponds to the cost of sending the data request to 
the data node, fetching the value of the stored data and sending back the stored value to 
the compute node. 
We need to consider two types of recurring costs after buying: $tRecMem$ - cost of 
computation at the compute node when data is in memory, and $tRecDisk$ - cost of 
computation at compute node after fetching data from disk. The costs of a compute request 
from compute node i to data node j are: 
\begin{flalign*}
&tCompute=max(tDisk_j,((s_k+s_p+s_{cv})/netBw_{ij}),tc_j )&\\
&tFetch=max(tDisk_j,((s_k+s_v)/netBw_{ij})) &\\
&tRecMem=tc_i &\\
&tRecDisk=max(tc_i,tDisk_i) &
\end{flalign*}

\noindent where $netBw_{ij}$ is the effective network bandwidth between the compute node 
$i$ and the data node $j$, which is computed during initial setup as described in 
Section~\ref{sec:cost}. The first component of $tCompute$ and $tFetch$ is the disk cost, 
while the second component takes into account the network cost.  
Since we use asynchronous calls, multiple invocations of the function $f$($k_i$,$p_i$) 
run concurrently and the disk and network access of these overlap with each other. 
The higher of the disk and network costs is the bottleneck of the system. Therefore, we 
take the maximum of the two costs. Similarly, $tRecDisk$ is the maximum of the CPU cost 
and the disk cost.

We assume that $tData$ is more than $tCompute$.
Clearly, if $tData \leq tCompute$ the decision would be to always issue data requests. 
Since the costs are key specific, the compute node would not be able to make the decision 
between compute and data requests until it has the cost computation parameters 
corresponding to the key. 
Therefore, the first request for a key is 
always sent as a compute request. The data node can choose to perform the function 
computation or to send data back. In either case, it sends the parameters for cost 
computation back to the compute node for future use. 

In order to apply  ski-rental based caching, we need to keep track of the number of times 
function calls are made for each key. Since the number of keys may be very large it may 
not be possible keep exact count for all keys.
As described in \cite{stream:frequent} there are several existing techniques to keep 
track of the most frequent values in a stream. We maintain the count of most frequent 
keys in buckets of hashmap using the Lossy Counting algorithm described in 
\cite{frequent:manku}.

\section{Balancing Computations}
\label{sec:load_balance}

Since the compute nodes do not have any stored state (other than transiently cached data 
items), each incoming tuple can be sent to any compute node. 
Hence, load balancing among compute nodes can be easily done by distributing the incoming 
stored or streaming tuples evenly across compute nodes. Load balancing among data nodes 
can be done if it is supported by the data store. For example, HBase has a balancer that 
balances the number of regions on different nodes. In this paper, we consider load 
balancing between the compute nodes and data nodes. This is particularly important 
in settings where the compute and data nodes are separate.

In the model we have described so far, a compute node always sends compute requests to 
data nodes if the key is not accessed frequently enough; if the key is accessed 
frequently its value is cached and the function is executed at the compute 
node.\footnote{ 
With moderate skew, it always makes sense to compute functions locally if the data is 
already cached. In case of very high skew along with high cost of function computation, 
this may result in compute nodes that are heavily loaded while data nodes may be less 
loaded. Our experimental results show that this happens only under very high skew and 
high compute costs. 
Extensions to make decisions on offloading computation to data nodes
for cached data items are a topic of future work.  
} This may cause a higher load at the data node as compared to compute nodes. We now 
describe how to balance the load between compute and data nodes.

To balance the load, a data node could choose to perform function execution for some 
requests, and for other requests return stored values to the compute node; in the latter 
case, the function computation would then be done at the compute node.
This choice is made based on the load at the data node and the load at the compute 
node.\footnote{An alternative would be to make this decision at the compute node;
however, this would require a message round trip to the data node 
to find its current load, which our approach avoids. Also note that centralized decision 
making is not feasible in our setting since it would not scale well.} 
Regardless of this choice, disk access cost will be incurred at the 
data node. To balance the load, we therefore consider only the network and CPU costs. 
We give a brief sketch on how the network and CPU loads at compute and data 
nodes are estimated. Details on how to compute these are discussed in 
\paper{\cite{throughputOpt:arxiv}.}\fullversion{Appendix~\ref{sec:loadComp}.}

Along with (a batch of) requests, the compute node also sends some statistical 
information to the data node. The statistics sent by a compute node include the 
number of requests pending to be computed locally, number of compute requests 
sent to all data nodes, average CPU time taken to compute a function, among others. The 
data node uses  these statistics from the compute node, as well as similar local load 
statistics, to estimate load at the compute and data node.

Formally, for each batch of data requests 
containing $b$ requests from a compute node $i$ to a data node $j$, to balance the load, 
the data node may choose to compute $d$ requests at the data node itself and 
send the remaining $(b-d)$ computations back to the compute node.

The CPU load at compute node $i$ (as a function of the number of requests $d$ from the 
batch that are compute at data node $j$) $compCPU_i(d)$, can be computed at a data node 
$j$ based on 4 components (1) the number of pending computations to be performed at $i$, 
(2) the {\em estimated} number of computations that are returned from the data 
nodes other than $j$ (these estimates are based on recent history), (3) the number of 
computations that are to be returned from $j$ to $i$ from previous requests pending at 
$j$ and (4) the number of requests, for the current  batch, that are to be computed at 
$i$ i.e. $b-d$. 

Similarly, the network load, $compNet_i(d)$ may be computed based on (1) the number of 
pending data and compute requests to be sent from compute node $i$ to data nodes, (2) the 
number of pending responses to data requests sent from $i$, (3) the {\em estimated} 
number of computed and uncomputed responses to compute requests made by $i$ to data nodes 
other than $j$, (4) the number of computed and uncomputed responses to compute requests 
to $j$ for previous requests, and (5) the number of computed and uncomputed responses for 
the current batch of requests ($d$ and $b-d$ respectively).

The CPU load at data node $j$, $dataCPU_j(d)$ may be computed based on the number of 
pending computations (from all compute nodes) to be computed at data node $j$ from 
previous batches, and the number of computations from the current batch of requests 
($d$). The network load at $j$, $dataNet_j(d)$ may be computed based on the 
amount of data that is to be sent for data requests and computed/uncomputed values for 
compute requests from all compute nodes, for previous batches as well as the current 
batch.

Since the computations at the compute node and the data node can go in parallel and the 
network transfer can go on concurrently with other computations at the compute and data 
nodes, the completion time for the batch is the maximum of the CPU and network time at 
the compute node and data node, i.e. $max($$compCPU_i(d)$, $compNet_i(d)$, 
$dataCPU_j(d)$, $dataNet_j(d))$. 
In order to get optimal throughput, the data node needs to minimize the completion time 
by selecting the optimal number of tuples $d$ to process at the data node.
We use gradient descent to compute the value of $d$ such that the cost is minimized. 
As shown in \paper{\cite{throughputOpt:arxiv}}\fullversion{Appendix\ref{sec:loadComp}} 
all costs are linear in $d$. This value needs to be computed for each batch of requests.
Hence, gradient descent is a cheap heuristic to compute $d$ even though it does not 
guarantee the global minimum. Note that data node $j$ makes the decision without any 
knowledge of the global load across all nodes, which allows this approach to scale.

We note that our approach also does some load balancing across compute nodes and across 
data nodes. Data nodes send back fewer computations to 
the compute nodes with high load, and send back more computations to compute nodes with 
lower loads, thus balancing the load between compute nodes. Similarly, data nodes with 
higher CPU loads send back more computations to the compute nodes and data nodes with 
less CPU load would send back fewer computations thereby mitigating the 
skew between data nodes. A similar effect happens with network load as well.

\section{Multiple Joins}
\label{sec:multiple}
In our discussion so far, we have focused on the join between the input streaming or 
stored data with only one stored relation. In general, the input stream/stored data could 
be joined with multiple stored relations, with each join possibly requiring some 
computation on the join result. 

Our approach can be easily extended to accommodate multiple joins without adversely 
affecting performance. In our framework, the compute node gets the result of the join.
Each join result is fed as the input of the next join (similar to left deep join 
plans) in a pipelined way to compute multiple joins.  
Ski-rental and load balancing techniques for each join may be done as before. When 
computing load at each node (for the purpose of load balancing) we consider 
the combined load from computing all the joins both at compute nodes and data nodes.

An alternate to our technique is to perform multiple joins using reduce-side joins in 
multiple stages; the output of each stage thus needs to be shuffled and partitioned using 
the join key for the next stage.  Since shuffling and partitioning are expensive 
operators, reduce side joins for multiple joins would be more expensive. Another issue 
with reduce side joins is in dealing with skew at each stage. It is difficult to 
estimate accurate statistics on join 
results. Approaches that use join statistics to mitigate skew by appropriate choices of 
reduce-side partitioning need the statistics before the join can start executing. This 
will prevent pipelining of join operators, thereby increasing the cost of execution.

Optimization of join order is an orthogonal problem and can be done either statically 
using standard query optimization techniques, or dynamically using  existing
techniques like STAIRs~\cite{stairs}, which dynamically decides between different join 
orders based on runtime statistics.

\section{Optimizing Calls to Data Store}
\label{sec:other_opt}

Requests to the data store are usually blocking. These blocking calls would lead to poor 
throughput since each process/thread would have only one request active at a time.
Sending one request at a time leads to poor resource utilization. 
Some data stores like HBase allow batch and asynchronous calls. If the data store does 
not allow for asynchronous calls, techniques from \cite{dbridge2015} can be used to 
implement batched asynchronous calls using multiple connections.

However, regardless of whether the data store provides asynchronous access calls to the 
data store, if the application is processing one input tuple (e.g. Hadoop, Storm) or one 
batch of tuples at a time (e.g. Spark Streaming, Trill)  it would block for the tuple 
(or batch) processing to finish. 
In order to efficiently use asynchronous calls, the application would need to be 
substantially changed. 
A cleaner approach is to issue prefetch requests, preferably in a separate function. We 
now discuss how to do this.

Prefetching and batching are supported for index nested loops joins in several databases 
like Oracle and Microsoft SQL server. However, our contribution is to provide an API and 
implementation for supporting it in Hadoop, Spark and Muppet.

\subsection{Prefetching}
\label{sec:prefetching}
\begin{figure}
	\begin{center}
		\includegraphics[width=0.48\textwidth,keepaspectratio=true]{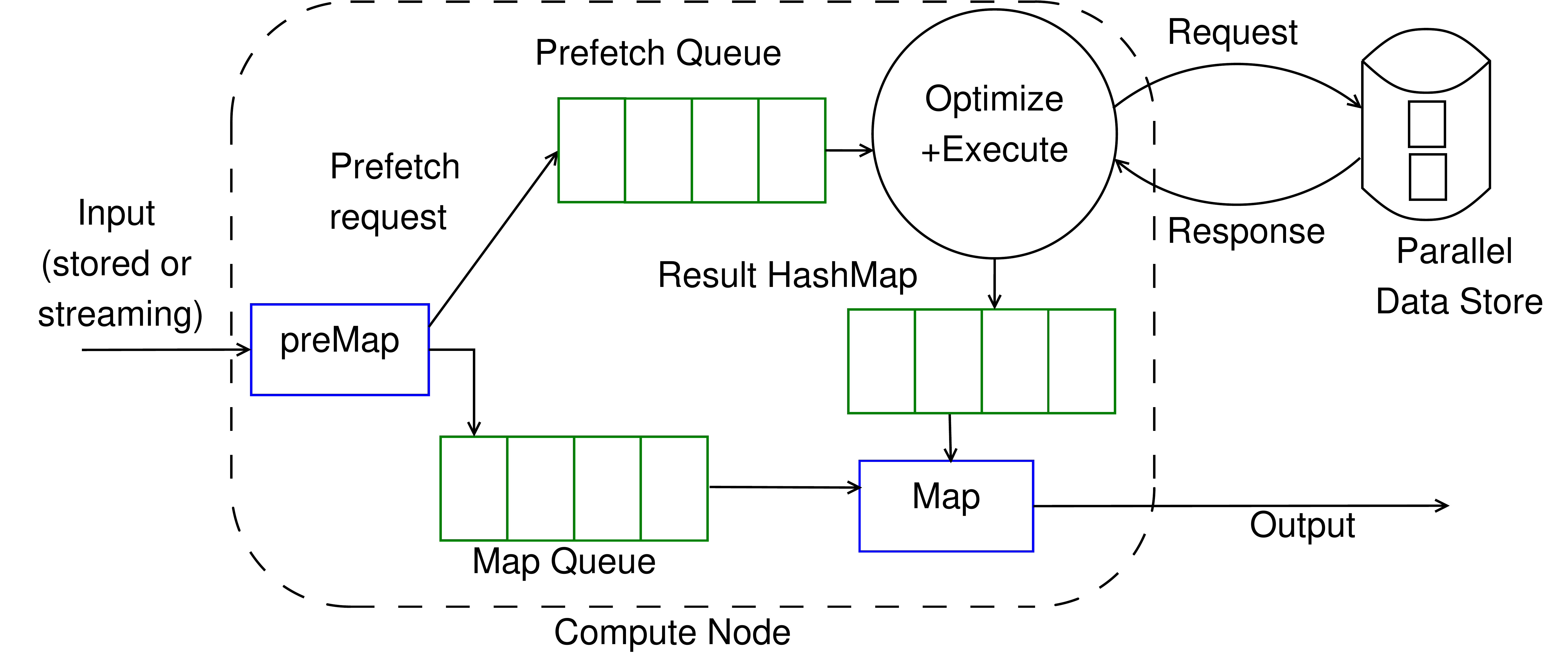}
		\caption{Invocation of preMap and Map}
		\label{fig:prefetch}
	\end{center}
\end{figure}

We extend the MapReduce API of Hadoop, map/flatMap functions of SparkRDD and  the 
MapUpdate API of Muppet by adding \smalltt{preMap} functions which  can be used to 
submit prefetch requests. The user can provide an implementation of the \smalltt{preMap} 
function to issue prefetch requests. 

A prefetch request returns immediately after taking steps to initiate fetch of 
results in the background. 
The driver function (the function that calls the map function at the 
compute node)  is modified so that the \smalltt{preMap} and the map functions 
run as separate threads. The \smalltt{preMap} function consumes data items from the 
source, issues prefetches and then adds the data items from input to a \smalltt{Map} 
queue as shown in Figure~\ref{fig:prefetch}. 
\revision{The compute node decides how to execute the function (at compute nodes or at 
data node) 
and sends appropriate requests to the data nodes. Once the function for a tuple is 
computed, it is added to a \smalltt{Result HashMap} from where the \smalltt{Map} function 
may read the computed value corresponding to the tuple read from the \smalltt{Map Queue}.}

In case of multiple joins, multiple such blocks as shown in Figure~\ref{fig:prefetch} can 
be pipelined. Instead of one $<$preMap,Map$>$ function pair, we create a series 
of $<$preMap,Map$>$ pairs, each to compute one join.

\subsection{Batching}

Sending data or compute requests individually may lead to poor performance. To improve 
performance, we batch multiple data fetch/prefetch or compute requests into one batch 
call.
Details of how to implement batching without modifications to the data store are 
provided in 
\paper{\cite{throughputOpt:arxiv}.} 
\fullversion{Appendix~\ref{app:compute_req_impl}.}

The batch size is currently decided statically, with batches kept large enough to
ensure per-request overheads are small relative to the actual cost of the request.
Extensions to dynamically determine batch size is a topic of future work.
Note that for streaming systems a large batch size is useful to improve throughput as 
shown in \cite{stream:adaptive:das}. However, batches must be kept small to keep the 
latency low. In order to keep the latency low, our framework allows applications to 
specify a maximum wait time. After the predefined amount of time has passed since the 
first data item was added to the queue, we send the batch irrespective of whether the 
batch is full or not. The waiting time to trigger a batch of requests can be adjusted 
depending on the latency requirements.

\section{Related Work}
\label{sec:relwork}

\paras{Stream relation join}
Prior work on optimization of stream-relation joins for 
non-distributed streaming systems includes
MeshJoin~\cite{meshjoin}, 
Semi-Streaming Index Join (SSIJ)~\cite{streamreljoin:icde11},
CacheJ\-oin~\cite{cachejoin}, and a technique proposed by
Derakhshan et al. in \cite{streamreljoin:cikm13}.

\eat{
In MeshJoin~\cite{meshjoin}, stream inputs are batched and the stored relation is loaded 
one page at a time in memory and the batched stream is joined with in-memory page.
Semi-Streaming Index Join (SSIJ)~\cite{streamreljoin:icde11} tries to maximize the 
throughput of stream-relation join by batching stream tuples loading pages of stored 
relation based on an optimal plan.
CacheJoin~\cite{cachejoin} keeps the most frequently accessed tuples of the stored 
relation in cache.
In CacheJoin the stream first looks up the cache to find a matching join tuple. If the 
lookup fails CacheJoin probes the relation from disk to perform the join. 
Derakhshan et al. in \cite{streamreljoin:cikm13} propose a new stream-relation join 
operator that relies on out of order process of data and batching stream tuples to 
improve the performance of stream-relation joins. 
}
However, none of these approaches consider the optimizations that we explore, 
such as prefetching and pushing computations to the data store. 
Since these techniques are not based on distributed streaming systems 
they also do not take into account load balancing.

\eat{Stream processing systems  like STREAM~\cite{stream:report} support join of streaming 
	data with stored relation. 
	Join of a stream with stored relations might be relatively slower since the stored relation
	may not fit in memory. }

\paras{Joins in distributed database systems}
Tian et al.~\cite{bigjoin} considers the problem of join of HDFS data with data in 
enterprise data warehouses, and consider several join techniques,
such as DB-Side join, HDFS-Side Broadcast join, HDFS-side repartition join,
and HDFS-side zigzag join, which are based on hash joins.
Indexed nested loops joins are not considered, which are 
key to handling streaming data, as well as in situations where
the stored data size is much larger than the other join input.
Further, they do not consider function executions.

\eat{
The focus is on using Bloom filters to minimize the amount of data to 
be transferred and 
perform hash based joins. Optimization of function execution is not considered.

We, on the other hand, rely on index nested loop join which can be done by using a HBase 
layer on top of HDFS. Hence, we can directly fetch the required data using the index to 
the data element.
}

\eat{
A number of techniques 
have been proposed to handle skew in joins for batch processing systems. 
DeWitt et al. \cite{skew:seshadri} use statistics on one relation to find heavy 
hitters, keep the heavy hitter tuples local and broadcast tuples from other relation that 
join with heavy hitters while \cite{flowjoin} use sampling to do the same. 
Another technique, as proposed in \cite{soumen:hipc},  is to pre-partition the stored 
data using cost estimates to mitigate skew not only in data distribution but skew in 
computation. The same partitioning function is used by the Mapper to distribute the input 
data to the appropriate reducer. 
These techniques rely on estimates that may not be accurate and require fixed threshold 
parameters to determine which keys are heavy hitters. 
Dynamic skew mitigation techniques like using a large number of virtual partitions 
relative to the number of nodes by adjusting partition boundaries.
SkewTune~\cite{skewtune} mitigates skew at runtime by dynamically re-allocating load.
However, they may not be efficient to mitigate skew when the skew is caused by a single 
key or a few keys.
}

Map-Side Index Nested Loop Join (MAPSIN join) \cite{mapsin} uses 
the indexing provided by HBase to perform map side joins on data 
stored in HBase. 
They do not perform any optimization when fetching data from other 
nodes in HBase, nor do they consider pushing computations to other 
nodes, or load balancing.

\paras{Skew reduction approaches}
DeWitt et al.~\cite{skew:seshadri} address the problem of skew in parallel hash join.
They use statistics to determine heavy hitters and broadcast heavy hitters from one of the
join inputs, to mitigate skew, while hash partitioning the other keys.
Flow-Join~\cite{flowjoin} uses approximate histograms to detect heavy-hitters
from an initial part of the input relations, and then uses the broadcast/hash partition
approach of \cite{skew:seshadri} to mitigate skew.  In particular, Flow-Join 
is optimized for very high bandwidth network interconnects.
Neither of these techniques considers function execution costs, nor
do they consider situations where the skewed keys change, as could happen
in a streaming system.  Furthermore, they depend on somewhat arbitrary
thresholds to determine which keys are heavy hitters, whereas our approach
based on ski-rental is more principled. \revision{However, our techniques have some 
overheads in terms of maintaining current statistics and caching. For in-memory systems 
with low latency network connectivity (like RDMA over InfiniBand network) and low CPU 
cost for function execution, the overhead of our techniques may increase the execution 
time, but they are useful when any of the above properties is not satisfied.}

SkewReduce mitigates skew by generating partitions using static optimization techniques 
while SkewTune performs dynamic repartition on detecting skew, to mitigate skew across 
Map and Reduce sites. Unlike these techniques, our techniques can mitigate skew where 
the skew is caused by heavy-hitters, since the computation for a single key can be
performed across multiple nodes.
Our system also performs many other optimizations like batching, prefetching and 
frequency based caching, which other systems do not consider.

\paras{Data access optimization}
The DBridge system~\cite{dbridge2015} provides APIs and implementations
for optimizing data accesses (e.g. using JDBC) from applications, by using batching, 
asynchronous invocations, and prefetching.  
Pyxis \cite{pyxis} partitions database applications into two parts, one to be run on 
the application server and the other at the database server, to minimize the amount of 
control and data flow for database calls.  
Both these approaches do not consider distributed systems, nor do they consider
dynamic runtime load balancing.  However, our implementation of batching/prefetching
uses the techniques described in \cite{dbridge2015}.

\paras{Load balancing}
Load balancing is a well-studied topic in the context of distributed and 
network systems, where the goal is to distribute the incoming requests
evenly among a number of nodes; see, e.g. ~\cite{loadbalalnce}. 
We instead look at balancing the function computation load between compute nodes and data nodes.
Our decisions are made considering a pair of nodes at a time, but are designed to take
into consideration load at other nodes as well when making the decision, to ensure
overall load balance across all compute and data nodes.  We are not aware of any 
other work which performs such optimization.

\section{Experiments}
\label{sec:expt}
For the purpose of evaluation of the efficiency of our techniques, we use our framework 
to optimize workloads on Hadoop YARN and Muppet, with HBase as the data store. 
We compare the performance of our techniques with that of existing skew mitigation 
techniques on an entity annotation application. For Spark, we compared the performance of 
our framework with SparkSQL on TPC-DS queries. We also compare the performance on 
synthetic workloads with different input skews for Hadoop and Muppet.

We use a cluster of 20 nodes to test the performance of our setup. Each node is equipped  
with two quad-core Xeon L5420 CPUs and 16 GB RAM. The amount of stored data stored in 
HBase was varied from 20GB to 200GB for different experiments.
We limited the cache size to 100 MB in memory to consider the scenario that memory cache 
is not enough to store all cached items. In practice, however,  data in the disk cache 
may actually be resident in memory as cached pages in the file-system buffer. Hence, 
reads from disk cache will incur file-system overhead, but may not incur actual disk 
access overhead, which can be very high for random IO on hard disks.
Thus, our numbers would more accurately match the cost of reads from an SSD, 
rather than from a hard disk.

\begin{figure*}
    \centering
    \begin{minipage}{0.32\textwidth}
            	\includegraphics[width=\textwidth,keepaspectratio=true]{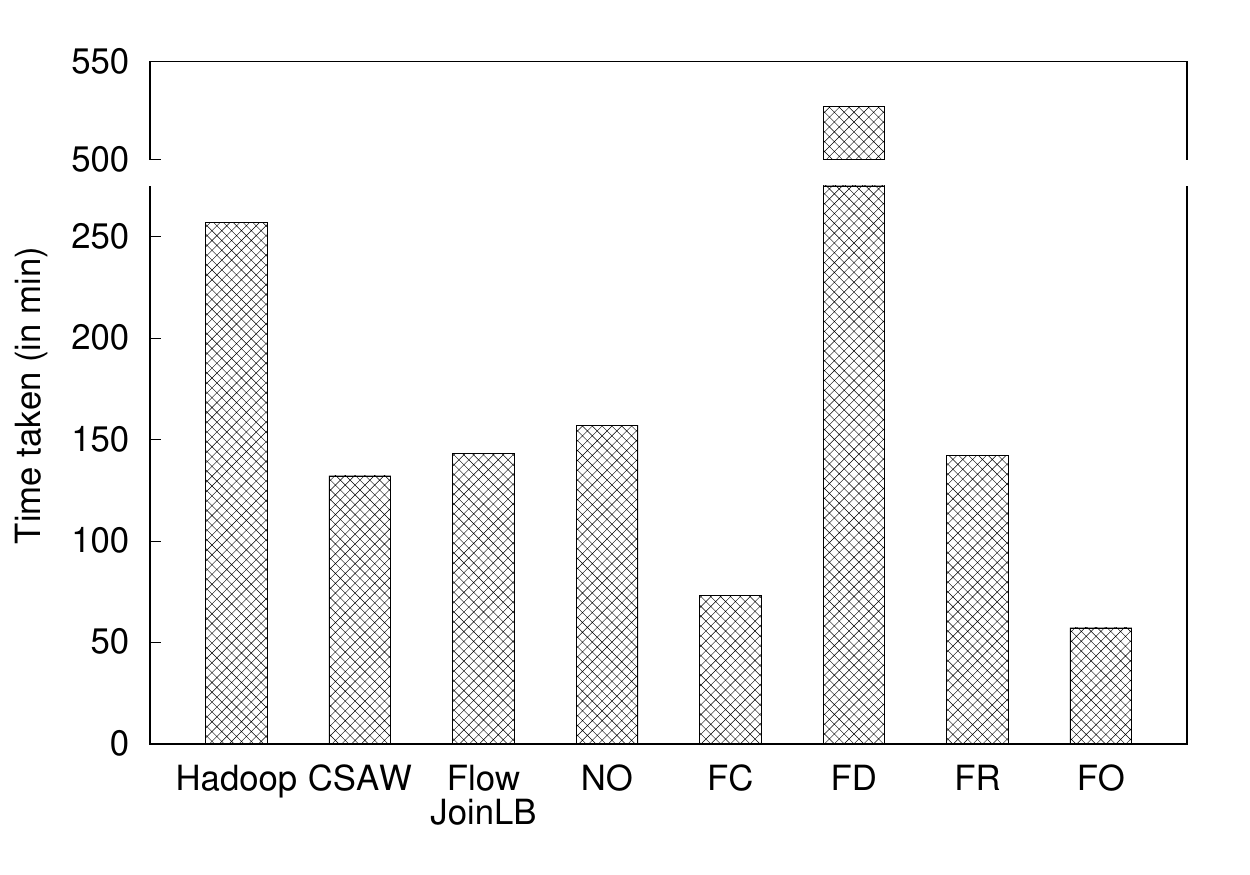}
            		\caption{ClueWeb dataset entity annotation using Hadoop MapReduce}
            		\label{fig:csawCompareYarn}
        \end{minipage}\hfill
    \begin{minipage}{0.32\textwidth}
        \centering
        \includegraphics[width=\textwidth,keepaspectratio=true]{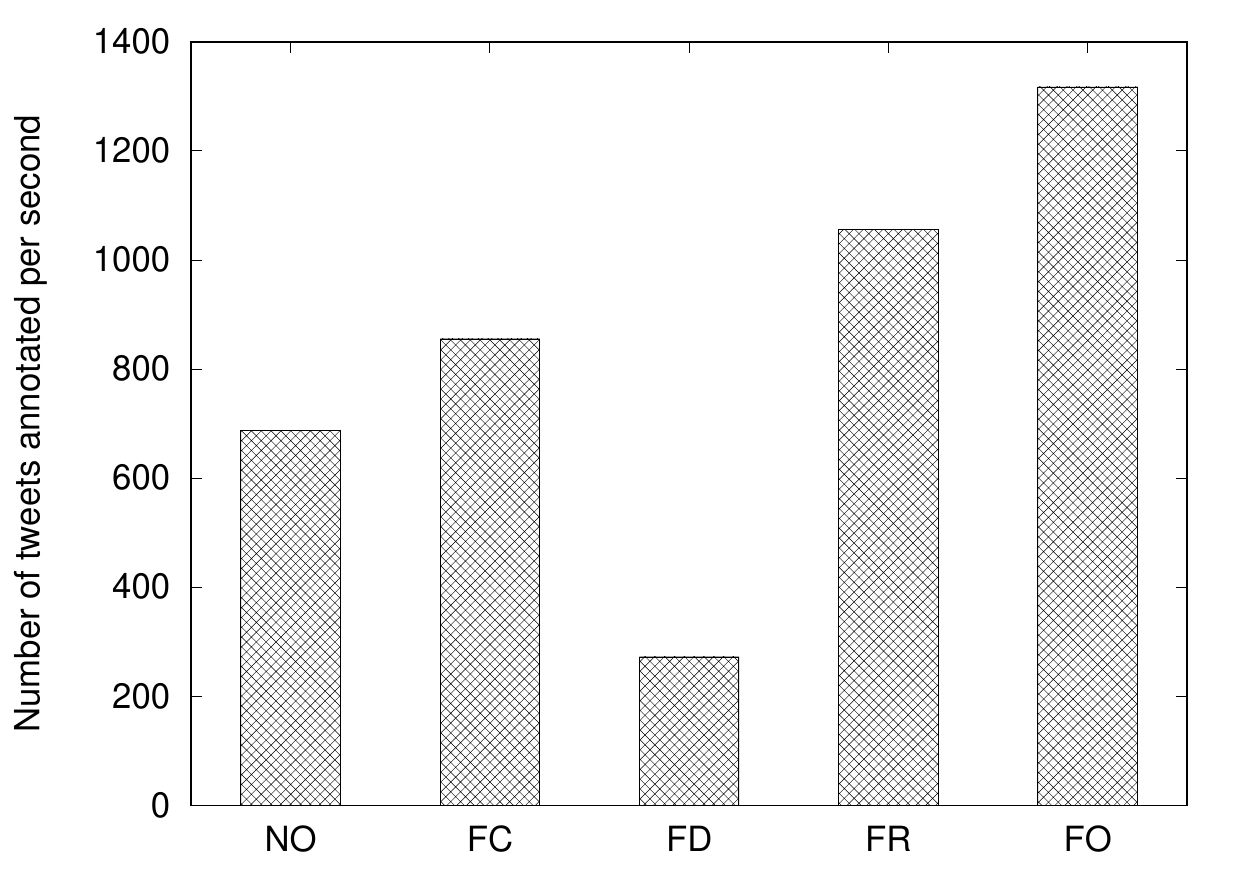}
        		\caption{Twitter entity annotation using Muppet}
        		\label{fig:csawCompareMuppet}
    \end{minipage}\hfill
    \begin{minipage}{0.32\textwidth}
            \centering
            \includegraphics[width=\textwidth,keepaspectratio=true]{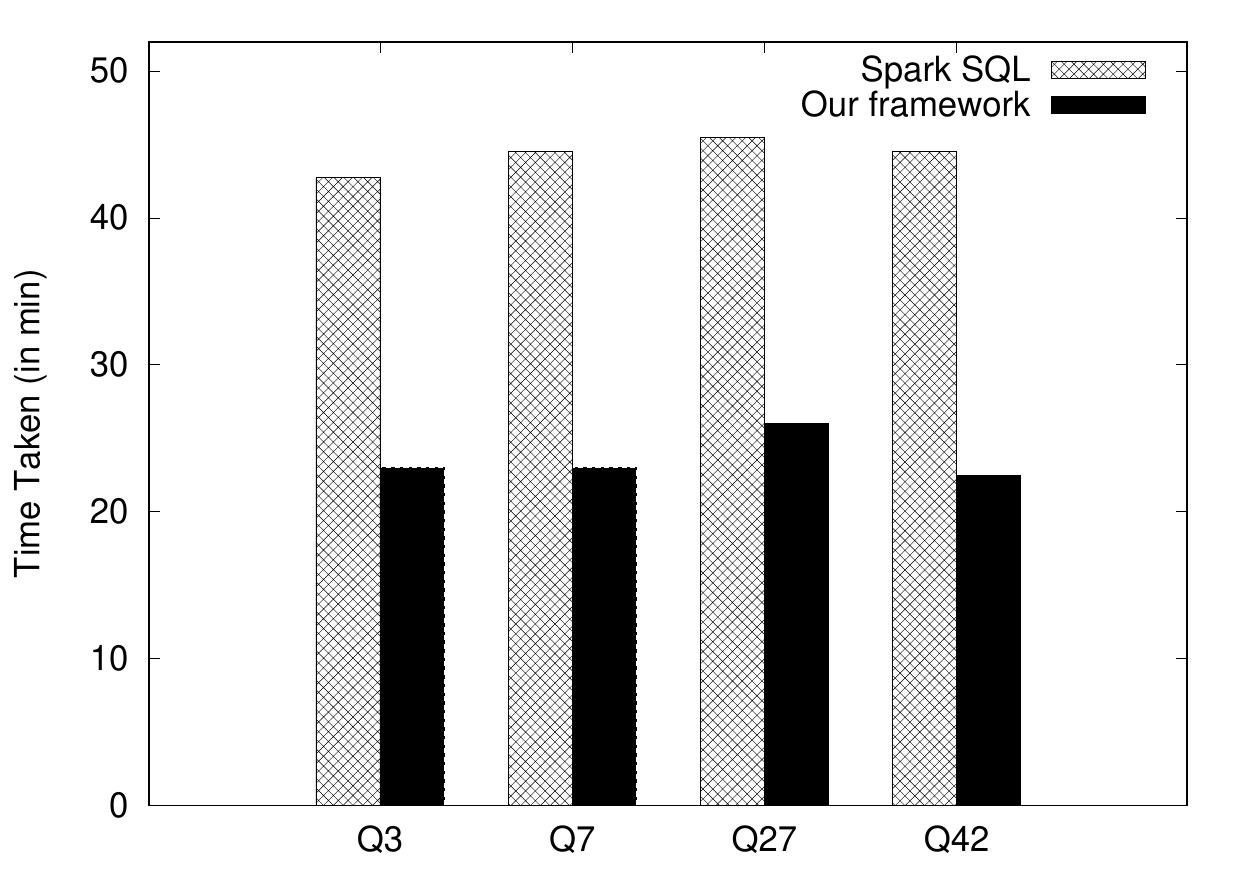}
            		\caption{Multiple join TPC-DS workload on Spark, SF=500}
            		\label{fig:multiple:spark}
        \end{minipage}\hfill
\end{figure*}

\subsection{Entity Annotation Workload}
For this experiment, we compare the performance of various techniques on an 
implementation of entity annotation using logistic regression models. 
The total data size of the models 28.7 GB with the largest being 284.7 MB and the 
smallest is just a few bytes.
Since this is highly CPU intensive even for a dataset of size 1 GB the basic 
MapReduce takes over 5 hours.

\subsubsection{Stored Data Performance}
In order to evaluate the performance of our optimizations, we compare across these 
options 
on Hadoop YARN.
\begin{itemize} [leftmargin=*] 
\setlength{\itemsep}{0pt}
\setlength{\parskip}{0.1pt}
 \setlength{\parsep}{0pt}
\item Hadoop: Basic Map Reduce in Hadoop with no skew mitigation techniques applied.

\item CSAW: The technique used in \cite{soumen:hipc} which estimates skew based on the 
frequency and the cost for annotation, and performs partitioning/replication accordingly,
as discussed earlier in Section~\ref{sec:entity}. 
\item FlowJoinLB: This technique uses statistics of the entire input data, and performs 
partitioning/replication as done in FlowJoin~\cite{flowjoin}. 
This technique provides a lower bound on the time taken by FlowJoin, which uses 
statistics based on a sample.

\item NO\fullversion{ - No Optimization}:  Map-side join, with data fetched using the 
default HBase APIs, and all function execution done at the compute nodes.  
None of our optimization techniques are used

\item FC\fullversion{ - Function execution at Compute nodes}:   Function execution is 
done only at the compute nodes.  Techniques of batching/prefetching are used to optimize 
data access, but no caching of data is done.

\item FD\fullversion{ - Function execution at Data nodes}: Function execution is done 
only at the
data nodes.  Techniques of batching/ prefetching are used to optimize data access. 
Data caching does not apply in this case.

\item FR\fullversion{ - Function execution with Random choice}:  The choice of 
compute/data request is made at random, with equal probability.  Techniques of batching/ 
prefetching are used to optimize data access, but no caching of data is done.

\item FO\fullversion{ - Function execution Optimized}:  All our optimizations are used,
including batching/ prefetching for optimizing data access, frequency based caching, 
and load balancing techniques.
\end{itemize}

We used a corpus of about 1GB/35,000 documents sampled from the ClueWeb09 
dataset {\cite{clueweb09}} and were able to annotate over 4.5 million 
entities.
To run entity annotations using MapReduce, CSAW and FlowJoinLB we used all 20 nodes on 
the cluster, while for the rest of the techniques we used 10 nodes as data nodes (for 
storing data in HBase) and 10 nodes as compute nodes (for running Hadoop).
Thus the total number of nodes used in the setup is same in all cases to provide a fair 
comparison.

We precompute statistics and cost estimates ahead of time for CSAW and FlowJoinLB and do 
not include the time taken. Our techniques do not need these statistics. However, all 
other overheads are included in the times reported.

In entity annotation, the next step is indexing which requires redistribution of 
tokens and entities; this incurs the same overheads for all techniques considered.
However, for applications that require the annotated results to be collected
with the document, our approach avoids further partitioning, since results are always
fetched to the compute nodes, where the document is processed.  In contrast, the other
techniques would incur an extra partitioning overhead in this case.

The total amount of time taken is shown in Figure~\ref{fig:csawCompareYarn}. A naive 
implementation using the Map Reduce framework causes high skew at reducers thereby 
creating stragglers. The straggler reducers increase the overall time taken. 
Similarly, for FD there was a lot of skew since some data nodes which contain heavy 
hitter keys or models for which classification is expensive, took much longer to complete.
CSAW and FlowJoinLB reduce skew, thereby improving the 
performance over the naive implementation. However, we still found some skew in the time 
taken by the reducers.

NO and FC perform the join and classification at Map side thereby reducing skew. However, 
they use only 10 nodes for computation. FR spreads computations across compute and data 
nodes but does it at random, thus leading to skew on some data nodes.
FO performs the best out of these techniques since the 
computation is spread uniformly across compute and data nodes due to runtime decisions 
made by our techniques. Note that FO takes less than half the time of CSAW, FlowJoinLB 
and FC takes 25\% more time than FO. FC can be significantly slower for other 
workloads such as Twitter annotation and synthetic workloads discussed later.

One interesting issue that we observed in our framework when using Hadoop was because 
of tasks restarts. Some map tasks straggled a little and could not finish as fast as 
others. The Hadoop framework restarted these map tasks on other nodes which led to 
extra function calls being pushed to the HBase store thereby reducing our performance 
slightly. However, this did not cause any material change to our result.

\subsubsection{Streaming Performance}
For this experiment, we compare the performance of entity annotation, in terms of the 
number of tweets processed, using Muppet on a 2 GB Tweet stream from June 2016. 
Since Muppet is a stream processing platform,  the MapReduce, CSAW and FlowJoinLB
techniques do not apply. The annotator was able to identify at least one entity for 
annotation in about 
50\% of the tweets. The number of tweets processed per second across the nodes is 
shown in Figure~\ref{fig:csawCompareMuppet}. FC was able to annotate more tweets than NO 
because it used batching and prefetching. FO performed significantly better since it was 
able to cache frequent models and balance computations between Muppet nodes and HBase 
nodes, and performs almost twice as well as NO. FD performed poorly due to skew.
We note that FO is only about 20\% better than FR; however that difference can be 
significantly more with skew as shown in the ClueWeb09 dataset and synthetic workload 
experiments later.

\begin{figure*}
 \centering
\begin{subfigure}{0.32\textwidth}
	\begin{center}
		\includegraphics[width=\textwidth,keepaspectratio=true]{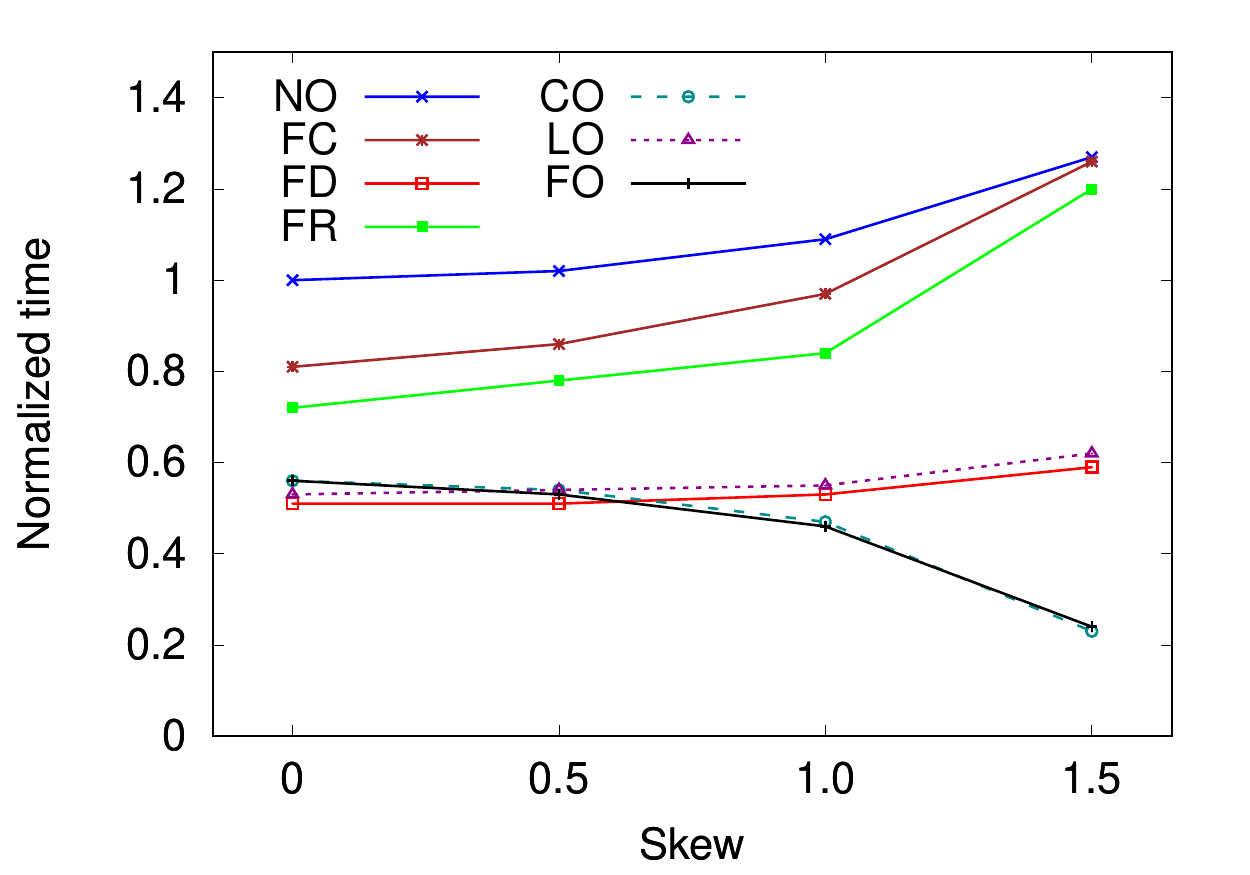}
		\caption{Data heavy workload}
		\label{fig:hadoopDh}
	\end{center}
\end{subfigure}
\hfill
\begin{subfigure}{0.32\textwidth}
	\begin{center}
		\includegraphics[width=\textwidth,keepaspectratio=true]{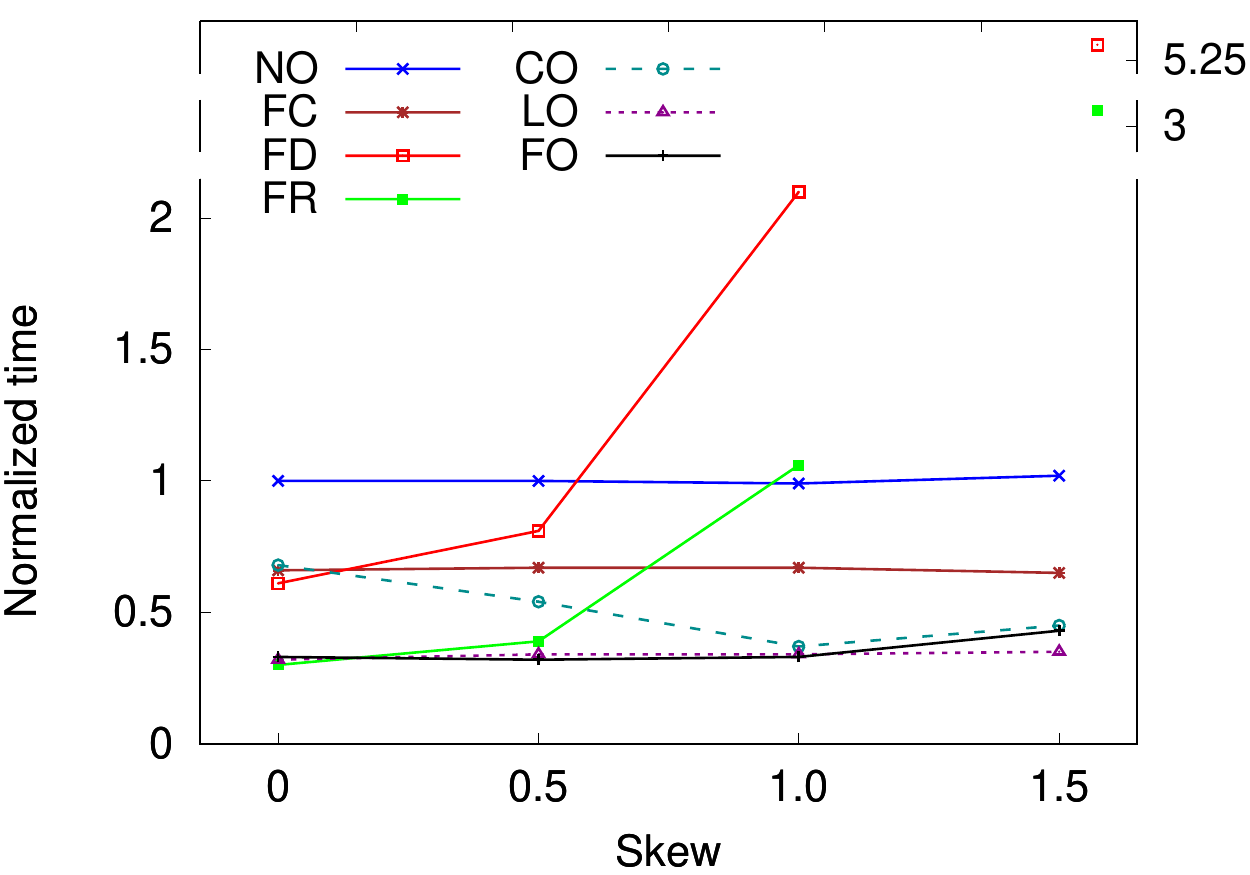}
		\caption{Compute heavy workload}
		\label{fig:hadoopCh}
	\end{center}
\end{subfigure}
\hfill
\begin{subfigure}{0.32\textwidth}
	\begin{center}
		\includegraphics[width=\textwidth,keepaspectratio=true]{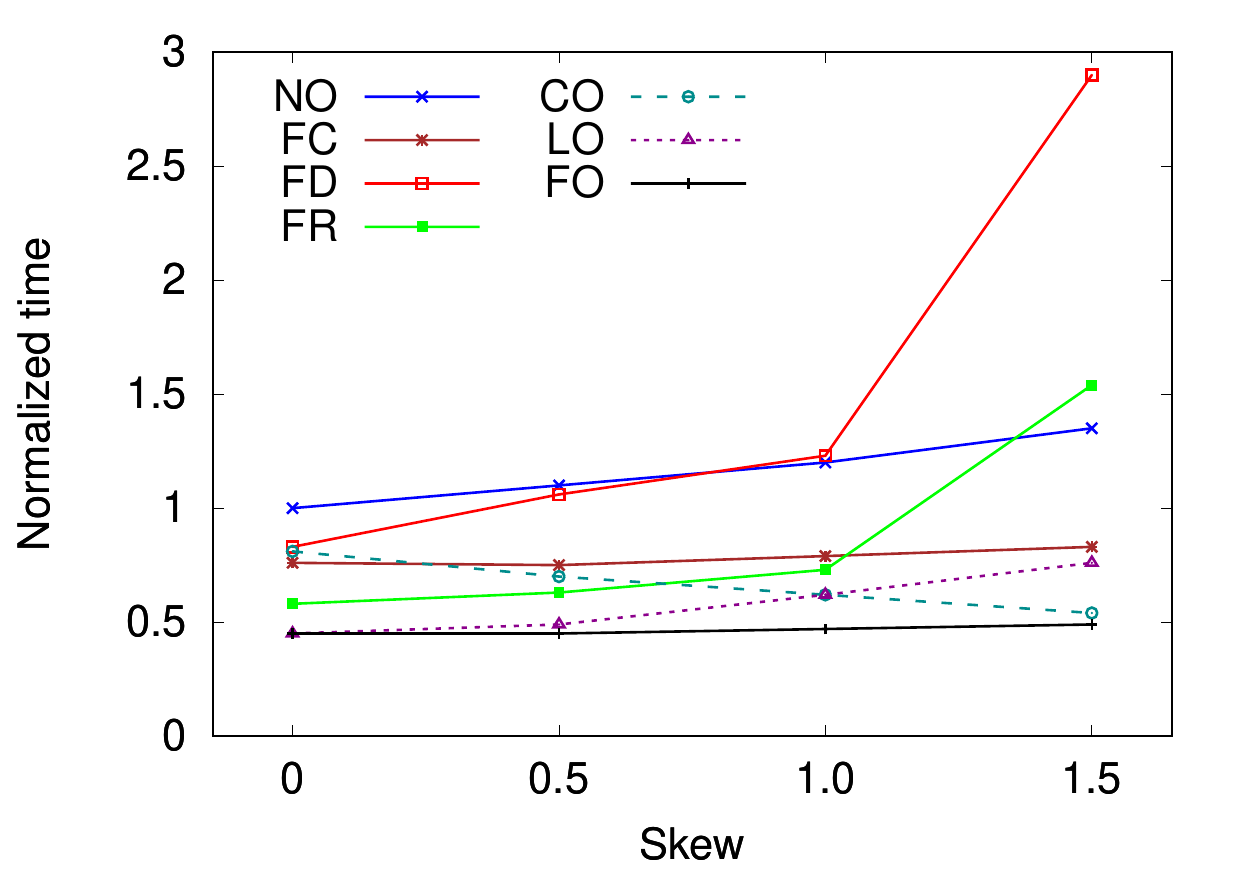}
		\caption{Data and compute heavy workload}
		\label{fig:hadoopCo}
	\end{center}
\end{subfigure}
\caption{Hadoop performance comparison for synthetic workload}
\label{fig:hadoopComp}
 \end{figure*}

\subsection{Multiple Joins and Spark Integration}
In order to evaluate our techniques for multiple joins, we used queries from the TPC-DS 
benchmark to evaluate our techniques on Spark. We used a scale factor of 500 to generate 
the TPC-DS dataset. We selected 4 queries that joined the \smalltt{store\_sales} relation 
with 2 to 4 other relations.

For Spark, we used a HDFS cluster of 20 nodes to store the data and ran queries using 
SparkSQL which internally uses the Catalyst optimization framework to optimize queries.
For our framework, we used 10 compute nodes to run Spark and 10 data nodes for HBase. The  
\smalltt{store\_sales} table was stored in HDFS across the 10 compute nodes and was read 
directly by Spark while other tables were stored in HBase. We used our extended Spark API 
to compute selections and joins (using the same join order as generated by 
Catalyst/SparkSQL) and then used SparkSQL for other operators (like aggregates, GROUP BY, 
ORDER BY, HAVING and LIMIT) on the join results.

The results of the experiment are shown in Figure~\ref{fig:multiple:spark}. 
For all queries, our framework performs better since it does not need to shuffle data for 
computing joins.

\subsection{Synthetic Workload}

For this experiment, we evaluate performance on the following synthetic workloads.
\begin{itemize}[leftmargin=*] 
\setlength{\itemsep}{0pt}
\setlength{\parskip}{0.1pt}
 \setlength{\parsep}{0pt}
	\item DH - Data Heavy workload: This workload computes a join and 
	projects attributes, returning only a small result. This workload is 
	heavy in terms of disk access and network but not on CPU. The size of data stored 
	in HBase for this workload was 200 GB with each data fetch being about 100 KB. The 
	total amount of data is more than the combined memory capacity of the data nodes 
	thereby ensuring that not all data items fit into memory.
	\item CH - Compute Heavy workload: This workload fetches only small amount of 
	data but performs some CPU heavy computations and simulates a compute heavy workload. 
	Each 
	computation takes about 100 ms on an average, while the total data size is 20 GB
	\item DCH - Data and Compute Heavy: This workload fetches large amounts 
	of data as well as performs CPU intensive computations. Each computation takes about 
	100 ms while the dataset is 200 GB.
\end{itemize}

\revision{ To distinguish the impact of caching from that of load balancing, we included 
the following in the performance evaluation.
 
 \begin{itemize}[leftmargin=*] 
 \setlength{\itemsep}{0pt}
 \setlength{\parskip}{0.1pt}
  \setlength{\parsep}{0pt}
 \item CO\fullversion{ - Ski-Rental optimization}:   Ski-Rental based caching is 
 used to find frequent items and cache them.  No load balancing is used. Techniques of  
 batching/prefetching are 
 used to optimize data access.
 \item LO\fullversion{ - Load Balancing}:   Load balancing techniques described in 
 Section~\ref{sec:load_balance} are used to balance the load between compute and data 
 nodes. No 
caching is done.  Techniques of  batching/prefetching are used to optimize data 
 access.
  
 \end{itemize}
 
 }
 
\subsubsection{Static Data Distribution}

To evaluate the performance of the synthetic workloads on Hadoop YARN we compare the 
performance, in terms of the amount of time taken, for NO, FC, FC, FR, CO, LO and FO 
described earlier. 
We use a Zipf function to generate the join keys with different skews. We varied the skew 
factor, z from 0 (uniform distribution) to 1.5 (highly skewed) with steps of 0.5. Note 
that there is no skew in the data stored in HBase since the key is a primary key and the 
size of each tuple is the same. We use 10 nodes as compute nodes 
and 10 nodes as data nodes.

Since our experiments are on a synthetic workload, the actual time is not relevant. 
Hence, we normalize the time taken by fixing the time taken for the NO technique at skew 
0 to be one unit of time and adjust the time taken for other techniques and skews 
appropriately. Across all workloads FC performs better than NO; the difference shows the 
benefits due to batching and prefetching optimizations.

The relative time taken for the data heavy workload is shown in 
Figure~\ref{fig:hadoopDh}.  For this workload, it would be beneficial to perform the join 
at the data node, since it would involve less network transfer cost. 
\revision{The performance of FO is marginally worse than FD at no skew because FO pays 
some overheads for cost estimation but in the end pushes compute requests to data nodes, 
just like FD. As the skew increases, FO caches the most frequent items and performs much 
better.
CO and LO show the impact of the caching and load balancing components of FO. 
In this case, CO is identical to FO because computation load is very low, so load 
balancing is not useful. LO performs slightly better at low skew because it has less 
overheads but at higher skew CO and FO perform better because of caching.}

Figure~\ref{fig:hadoopCh} shows the relative time taken for the compute heavy workload.  
At z=0, FR is able to evenly distribute the compute load between the 
compute and data nodes and hence performs very well. However, with increase in skew FR 
sends many compute requests to data nodes with skewed keys, and the performance falls. 
Similarly, for FD the time taken increases with increase in skew because some data nodes 
get very heavily loaded. 
At z=0, CO sends all computations to the data nodes and performs similar to FR.
  At higher skew CO is able to cache skewed values at compute nodes and 
offload some computations from the data nodes. However, LO and FO are able to better 
balance the computations between the compute and data nodes at all skews and hence 
outperform CO. 
At skew of z=1.5, the performance of FO decreases slightly since most of the computations 
are performed on cached items and our techniques perform the computation 
for cached items at compute nodes, leaving data nodes under utilized; LO, which does not 
cache data, is able to balance the load slightly better. Extending our load balancing 
techniques to detect and handle such execution skew is a topic of future work.

The relative time taken for the compute and data heavy workload is shown in 
Figure~\ref{fig:hadoopCo}. 
For this workload, FO outperforms FR even at z=0 because it balances between compute and 
data requests in a cost based manner, taking into account both the compute and data 
costs, while FR sends requests randomly. Similar to the compute heavy workload, as skew 
increases performance of FR reduces because it overloads some data nodes with too many 
requests. At low skew, CO is not able to cache values and hence sends all 
computations to the data nodes. However, with increase in skew CO is able to cache values 
for skewed keys and performs some computations at the compute nodes and hence its 
performance increases. 
LO does not do any caching and at high skew it needs to fetch a lot of data from data 
nodes that contain higher skewed values. Hence, its performance decreases with increase 
in skew. FO works well across all skew values.

\revision{We also ran this experiment on the Muppet stream processing system. Results are 
presented in \paper{\cite{throughputOpt:arxiv}.} 
\fullversion{Appendix~\ref{sec:implementation}.} The performance benefit of the 
optimizations for Muppet on these workloads is very similar to that of Hadoop. Across the 
board, FO performs best or close to the best. }

\subsubsection{Dynamic Data Distribution}

To evaluate our performance for changing distribution, we conducted another experiment 
with synthetic workloads where we dynamically changed the distribution. For each skew 
value, we changed the frequent keys 10 times during our experiment.

For the first set of experiments, we compared the performance of the earlier mentioned 
techniques on the dataset with changing distribution. We did not observe any 
noticeable difference and the results were similar to the one shown in 
Figure~\ref{fig:hadoopComp}. 

\begin{figure}
	\begin{center}
		\includegraphics[width=0.25\textwidth,keepaspectratio=true,angle=90]{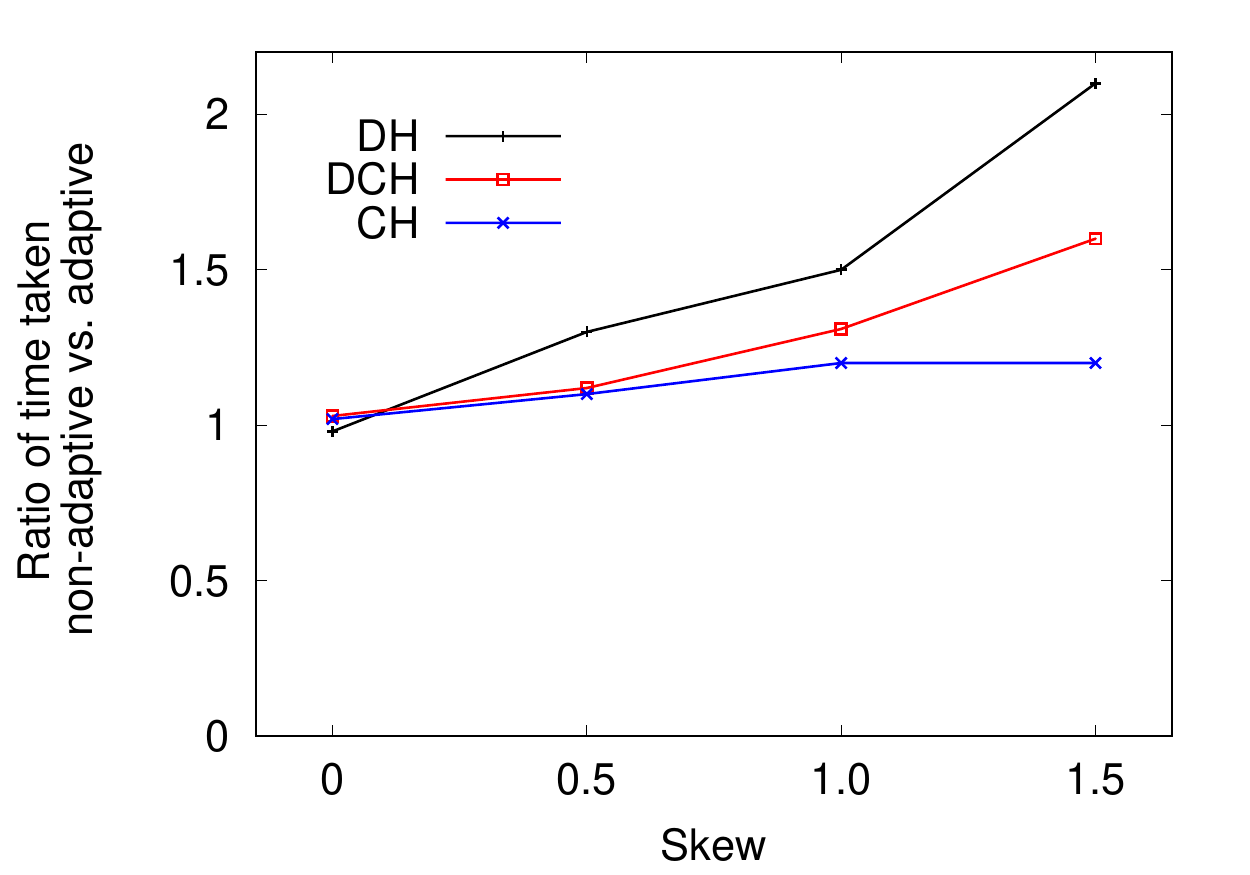}
		\caption{Adaptive vs non-adaptive optimization with load balancing}
		\label{fig:adaptive}
	\end{center}
\end{figure}

For the second set of experiments, we compare the performance of FO, which adapts 
continuously, with that of non-adaptive optimization. For 
non-adaptive optimization, ski-rental based caching decisions are made for only the first 
10\% of the tuples and the cache contents are not changed subsequently. Load balancing 
was performed as before.

Figure~\ref{fig:adaptive} shows the ratio of the time taken by the non-adaptive technique 
to that of the adaptive technique for the data heavy (DH), compute heavy (CH) and the 
data and compute heavy (DCH) workloads. When there is no skew (i.e., uniform 
distribution), the adaptive and non-adaptive techniques work equally well. For the 
compute heavy workload, CH the non-adaptive technique is able to balance the load among 
data and compute nodes and hence adaptive performs only slightly better. For DH and DCH, 
which benefit more from caching frequent values, the adaptive technique performs 
significantly better than the non-adaptive technique.

\section{Conclusion}

In this paper, we looked at techniques to optimize remote data access and function 
invocations by optimizing the location of function execution using ski-rental based 
caching and load-balancing techniques, along with prefetching and batching optimizations 
for applications running on parallel data frameworks. 
Our performance results show significant benefits in terms of throughput improvement
across different frameworks and workloads.
Our techniques can also be used for optimization of joins, with optional
UDFs, in parallel databases.

Areas of future work include dynamic choice of batch size 
and batch timeout taking latency into account, handling user defined functions with side 
effects and elastically increasing or decreasing compute nodes based on load. Another 
area of future work is to extend the Catalyst optimizer of SparkSQL to use our join 
technique when appropriate.

\section*{Acknowledgment}
The work of Bikash Chandra is supported by a fellowship from Tata Consultancy Services.
\bibliographystyle{abbrv}
 \bibliography{references}
\fullversion{

\renewcommand\thesection{A\arabic{section}}
\appendix

\section{Other Application Domains}
\label{app:applications}

CloudBurst~\cite{cloudburst} aligns a (large) set of genome sequence reads (which are typically
small) with a reference genome sequence, to find locations in the reference sequence that approximately
match each read.  CloudBurst is implemented using MapReduce.    
One map function extracts n-grams from the reads, and outputs (n-gram, string) pairs,
with the n-gram as the key for the reduce phase.  
A similar map function extracts n-grams from the reference sequence, and outputs (n-gram, string)
pairs.  The reducer for a particular n-gram matches each read with the reference sequence string
at the matching location using approximate matching algorithms.  
As pointed in \cite{hadoopskew}, the basic MapReduce implementation leads to skew, with one of 
the important causes being a variance in cost of user defined operations (UDOs) executed at the reducers
(UDOs correspond to UDFs in our framework).  
The SkewTune technique of \cite{hadoopskew} detects skew and repartitions tasks assigned to straggler
reducers, to mitigate skew.    

The reduce function in CloudBurst basically performs a join, followed by a UDF computation.
Our framework could be used to handle this problem as follows.  The reads are partitioned
amongst compute nodes.  The n-grams from the reference sequence are computed and indexed with
the n-grams as keys, and the strings around the n-gram location as values.  The map function
extracts n-grams from the reads, and for each n-gram, fetches matching reference strings from 
the datastore.  The approximate matching is done as a UDF.
Note that if the join is done at the reducer, all reads with a particular n-gram would go 
to a single reducer; whereas in our case the join can be performed at the map-side,
and these n-grams thus get distributed across multiple compute nodes, evening out the UDF 
load. Note that unlike SkewTune, our techniques are applicable to streaming systems also.

As mentioned in \cite{hadoopskew} there are many MapReduce applications that use UDOs and 
suffer
from skew.  We do not have details of the applications, but we believe our techniques would 
be applicable to at least some of them.

\section{Managing Memory \& Disk Caches}
\label{app:caching}

As discussed in Section~\ref{sec:ski:caching}, the \smalltt{condCacheInMemory} function 
is used to determine if a data item is to be cached in memory or disk. 
We describe this function in detail in this section.
We first consider the simpler case where all data items are the same size,
and then consider the general case where data items can be of different sizes.

We denote the memory cache as \smalltt{mCache}, and the disk cache as \smalltt{dCache} 
and their respective sizes be $s_{m}$ and $s_{d}$.

\subsection{Uniform data item caching}

The \smalltt{condCacheInMemory} for the case where data items are of uniform sizes is 
described in Algorithm~\ref{algo:uniformCache}. Given a new data item, this function 
checks if it is to be cached in memory or not. If the free space in cache is more than 
the size of the item, it is added to cache.  If not, the algorithm compares 
the benefit of the new data item and the item with the lowest benefit in 
\smalltt{mCache}. If the benefit of the new data item is higher, 
the item with the lowest benefit in \smalltt{mCache} is moved to disk cache, and
the current data item is cached in memory.
In both the above cases the function returns TRUE otherwise, it returns FALSE.

Note that we assume that the disk cache is large enough to cache all fetched items. In 
case the disk cache is full, items from disk cache with low benefit could be evicted 
from \smalltt{dCache} in order to accommodate items with a higher benefit. 
Also, data items that are moved to \smalltt{mCache} in the function 
\smalltt{condCacheInMemory} 
could be removed from \smalltt{dCache} to save space in the disk cache (although there
would be a cost to moving it back to disk in case it is evicted from memory).

\subsection{Non-uniform data item caching}
In case the size of data items is variable, we use Algorithm~\ref{algo:nonUniformCache} 
to check whether to put an item in \smalltt{mCache}. 
Unlike the fixed size case, removing one item from \smalltt{mCache} may not 
create sufficient free space to add a new item.
We, therefore, find $i$ items with the least priority such that eliminating these items 
would free up enough 
space to add the new item to \smalltt{mCache}. If the sum of priorities of the items 
to be evicted is less than the priority of the new data item, we return FALSE.
Note that since we chose the items in increasing order of benefit until there was enough 
space,
it is possible that some subset of them may actually suffice to create enough space
for the new item.  We, therefore, pick items with the most benefit that can be retained, 
while freeing up enough space for the new item. The remaining items are 
evicted to disk, and add the new item is added to \smalltt{mCache}. 

\begin{algorithm}
\renewcommand{\algorithmicrequire}{\textbf{Inputs:}}
    \renewcommand{\algorithmicensure}{\textbf{Output:}}
\begin{algorithmic}[1]
\REQUIRE k = key to be updated, v = value of item,  s = size of item
\IF{mCache.freeMem $\geq$ s}			  
   \STATE mCache.add(k,v)
   \RETURN TRUE 
\ELSIF{benefit(k) $>$ mCache.minBenefit}
  \STATE minPE $\leftarrow$ mCache.minBenefitElement
  \STATE mCache.remove(minPE)
  \STATE memCache.add(k,v)
  \IF{minP $\notin$ dCache}
	  \STATE dCache.add(minPE)
  \ENDIF
  \RETURN TRUE
\ELSE
  \RETURN FALSE
\ENDIF
\end{algorithmic}
\caption{: condCacheInMemory \\(with uniform item size)}
\label{algo:uniformCache}
\end{algorithm}

Note that after the above step, there may be some free space in \smalltt{mCache}. 
We do not actively pull elements from disk cache to fill up the free space. 
Instead, we fill the free space lazily when other items are accessed.

Note that as for the fixed item size case, we assume that disk cache is unlimited. 
In case the disk cache size is limited, items from disk cache with low benefit to
size ratio could be evicted from disk cache in order to accommodate items with 
higher benefit to size ratio.

\begin{algorithm}
\renewcommand{\algorithmicrequire}{\textbf{Inputs:}}
\renewcommand{\algorithmicensure}{\textbf{Output:}}
\begin{algorithmic}[1]
\REQUIRE k = key to be updated, v = value of item,  s = size of item

\IF{mCache.freeMem $\geq$ s}			  
   \STATE mCache.add(k,v)
   \RETURN TRUE 
\ELSE 

	\STATE prelimList $\leftarrow$ i items with least benefits $\mid$ $\sum$size + 
	mCache.freeMem$>$ s
	\STATE freeMem  $\leftarrow$ mCache.freeMem+ $\sum_{i}$ preimList(i)
	\IF{$benefit(k) \geq \sum$ benefit(prelimList)}
		\STATE keepList $\leftarrow$ items in prelimList with most benefit $\mid$ 
		size(keepList) $\leq$ freeMem$-$s
		\STATE removeList $\leftarrow$ prelimList - keepList
	
		\FOR{item $\in$ removeList}
			\STATE mCache.remove(item)
			\IF{item $\notin$ dCache}
				  \STATE dCache.add(item)
			  \ENDIF 
		\ENDFOR
		\STATE mCache.add(k,v)
		\RETURN TRUE
	\ELSE
		\RETURN FALSE
	\ENDIF
\ENDIF
\end{algorithmic}
\caption{: condCacheInMemory \\ \hfill{(with variable size items)}}
\label{algo:nonUniformCache}
\end{algorithm}

\section{Load computation}
\label{sec:loadComp}

In Section~\ref{sec:load_balance}, we discussed how on receiving a request batch from a 
compute node, a data node balances the load between itself and the compute node by 
estimating its own load and the load at compute and then deciding 
to compute only a fraction of requests from the batch locally. We now discuss how a data 
node estimates load, based on the statistics sent to it from compute nodes and statistics 
available locally. 

Consider a batch of $b$ requests sent from the compute node $i$ to the 
data node $j$. The data node chooses to compute $d$ requests at the data node 
itself and send $b-d$  computations back at the compute node. 

The load at compute node $i$ at a point in time is estimated based on the 
following parameters, which are measured continuously at runtime. (Parameters sent from 
the compute node are marked with superscript $c$ while those computed at the data node 
are marked with superscript $d$.)

\begin{itemize} \setlength{\itemsep}{0pt}
\item $lc^c_i$: number of pending local computations (based on fetched values) at 
compute node $i$ 
\item $nd^c_i$: number of pending data requests to be sent from compute node $i$ 
\item $nc^c_i$: number of pending compute requests to be sent from compute node $i$ 
\item $ndr^c_i$: number of pending responses to data requests sent from compute node $i$
\item $\overline{nr}^c_{ij}$:\eat{= $\sum_{k\neq j}{nr^c_{ij}}$:} total number of pending 
compute requests at compute node $i$ across data nodes other than $j$
\item $\overline{r}^c_{ij}$:\eat{=$\sum_{k\neq j}{r^c_{ij}}$} number of pending 
compute requests at compute node $i$ 
expected to be computed at the data nodes other than $j$ (this is computed based on 
recent history)

\item $nr^d_{ij}$: number of pending compute requests sent to data node $j$ from compute 
node $i$
\item $r^d_{ij}$: number of pending compute requests sent to the data node $j$ from the 
compute node $i$ that are to be computed at the $j$

\eat{\item $fc^c_{ij}$: the fraction of compute requests each data node $j$ has actually 
computed in the recent past 
\item $r^c_{ij}$ = $fc^c_{ij}*nr_{ij}$: expected number of pending compute requests (sent 
by the compute node $i$) to be computed at data node $j$}
\end{itemize}

The load at a data node $j$ can be estimated using the following parameters.

\begin{itemize}\setlength{\itemsep}{0pt}
\item $nd^c_j$: number of data requests pending at data node $j$ from all compute nodes
\item $ndr^d_j$: number of pending data request responses to be send from data node $j$
\item $nr^d_{j}$\eat{= $\sum_i {nr^d_{ij}}$}: total number of pending compute requests 
(from all compute nodes) at data node $j$ (note that some of these may be sent back to 
the compute nodes) 
\item $r^d_{j}$\eat{ = $\sum_i{r^d_{ij}}$}: number of pending compute requests (from all 
compute nodes) to be computed at the data node 
\item $nr_{ij}^d$: the number of pending compute requests at data node $j$ from compute 
node $i$
\item  $r_{ij}^d$\eat{= $fc^d_{ij}*nr^d_{ij}$}: number of pending compute requests from 
the compute node $i$ to be computed at the data node $j$

\eat{\item $fc_{ij}^d$: the fraction of compute requests from compute node $i$ that data 
node $j$ had computed in the recent past}
\end{itemize}

\eat{Note that the $nr_{ij}^c$ and $nr_{ij}^d$ denote basically the same value but as 
computed at the compute node $i$ and data node $j$ respectively. Similarly, for 
$fc_{ij}^d$ and $fc_{ij}^c$, as well as for $r_{ij}^c$ and $r_{ij}^d$. As described in 
Table~\ref{tab:cost}, $tc_i$ and $tc_j$ are the average time taken to 
compute the function at nodes $i$ and $j$  respectively.
}

Let the time taken to compute the function at compute node be $tc^c$ while the time taken 
to compute a function at data node be $tc^d$. 
Let the size of the key be $s_k$, the size of the parameters be $s_p$, the size of the 
store value be $s_v$ and the size of the computed value be $s_{cv}$.
The load is estimated in terms of CPU and network as shown below.

The CPU load at compute node $i$ is the time taken for 
computation of (1) the number of pending computations to be performed at $i$ ($lc^c_i$), 
(2) the {\em estimated} number of computations that are returned from the data 
nodes other than $j$ (these estimates are based on recent history) 
($\overline{nr}^c_{ij}-  \overline{r}^c_{ij}$),  (3) the number of 
computations that are to be returned from $j$ to $i$ from previous requests pending at 
$j$ ($nr^d_{ij}- r^d_{ij}$) and (4) the number of requests, for the current  batch, that 
are to be computed at $i$, i.e. $b-d$. Hence, the CPU load at compute node $i$ can be 
estimated as
\begin{equation*}
\begin{aligned}
compCPU(d)	= tc^c*lc^c_i + tc^d* (\overline{nr}^c_{ij}-  \overline{r}^c_{ij}) \\
+ tc^d* (nr^d_{ij}- r^d_{ij}) +tc^d*(b-d)
\end{aligned}
\end{equation*}

Similarly the CPU load at data node $j$ can be estimated as 
\begin{equation*}
dataCPU(d) =	tc^d* r^d_{j} + tc^d * d
\end{equation*}

The time taken for network communication for a node is $networkLoad/netBw$.
The network load at the compute node $i$ is the sum (1) the number of 
pending data and compute requests to be sent from compute node $i$ to data nodes 
($nd^c_i$ and $nc^c_i$ respectively), (2) the number of pending responses to data 
requests sent from $i$ ($ndr^c_i$), (3) the {\em estimated} number of computed and 
uncomputed responses to compute requests made by $i$ to data nodes other than $j$ 
($\overline{r}^c_{ij}$ and $\overline{nr}^c_{ij}-\overline{r}^c_{ij}$  
respectively), (4) the number of computed and uncomputed responses to compute requests to 
$j$ for previous requests (${r}^d_{ij}$ and ${nr}^d_{ij}-{r}^d_{ij}$ respectively), and 
(5) the number of computed and uncomputed responses for the current batch 
of requests ($d$ and $b-d$ respectively).
 Thus the network load at the compute node $i$, is a function can be computed as 

$compNet(d)=$
\begin{equation*}
\frac{\splitdfrac{
	nd^c_i*(s_k+s_v)+nc^c_i*(s_k+s_p) + ndr^c_j*s_v}
	{\splitdfrac{+(\overline{nr}^c_{ij}-\overline{r}^c_{ij})*s_v	+ 
	\overline{r}^c_{ij}*s_{cv}}
	{\splitfrac{+({nr}^d_{ij}-{r}^d_{ij})*s_v	+{r}^d_{ij}*s_{cv}} 
	{+d*s_{cv}+(b-d)*s_v}}}}
{netBw_i}
\end{equation*}

\eat{
\begin{equation*}
\begin{aligned}
compNet(d)=[nd^c_i*(s_k+s_v)+nc^c_i*(s_k+s_p) + ndr^c_j*s_v\\
	+(\overline{nr}^c_{ij}-\overline{r}^c_{ij})*s_v	+ 
	\overline{r}^c_{ij}*s_{cv}
	+({nr}^d_{ij}-{r}^d_{ij})*s_v \\+ {r}^d_{ij}*s_{cv} 
		+d*s_{cv}+(b-d)*s_v]/
netBw_i
\end{aligned}
\end{equation*}}
Similarly, the network load at data node $j$ is a function of $r_j$,  

$dataNet(d)=$
\begin{equation*}
\dfrac{\splitdfrac{nd^c_j*(s_k+s_v)+ ndr^d_j*s_v+nr^d_{j}*(s_k+s_p) } 
		{\splitfrac{+(nr^d_{j}-r^d_{j})*s_v+ r^d_{j}*s_{cv}} 
		{+d*s_{cv}+(b-d)*s_v}} 
		}		
{netBw_j} 
\end{equation*}

As described in Section~\ref{sec:load_balance} we need to choose $d$ that minimizes the 
maximum of the four loads described above. All the functions are linear in d. The choice 
of $d$ needs to be made for each batch of compute request. We use gradient descent as a 
cheap heuristic to compute d even though it does not guarantee a global minimum and may 
get stuck at a local minimum. The gradient descent  initially starts with a random point, 
$\bar{d}$ between $0$ and $b$ which gives us an initial value for the max function. We 
then iteratively follow the decreasing slope till we get a minimum.

\section{Implementation}
\label{sec:implementation}

In this section, we describe our cache implementation as well as our implementation of 
preMap for Hadoop MapReduce and Muppet stream processing framework. We also describe how 
we use batch compute 
requests for HBase and measure network bandwidth.

\begin{figure}
\begin{scriptsize}
\begin{verbatim}
preMap(docId,document) {
  for each spot in document.getSpots() {
   spotContext = 
        getContextRecord(spot, document)
   submitComp(f,spotContext.key,spotContext.value) 
  }
  mapQueue.add([docId,document])
}

map(docId,document)
  for each spot in document.getSpots() {
    spotContext = getContextRecord(spot, document)  
    annotatedValues = 
     fetchComp(f,spotContext.key,spotContext.value)    
    context.write(annotatedValues)
  }
}
		
f(key, params){
  model = getModel(key)
  annotatedValues = classifyRecord(params, model)
  return annotatedValues
}
\end{verbatim}
\end{scriptsize}
\vspace{-0.3cm}
\caption{Entity annotation using prefetching}
\label{fig:opt}
\end{figure}

\begin{figure*}
\centering
\begin{subfigure}{0.32\textwidth}
	\begin{center}
		\includegraphics[width=\textwidth,keepaspectratio=true]{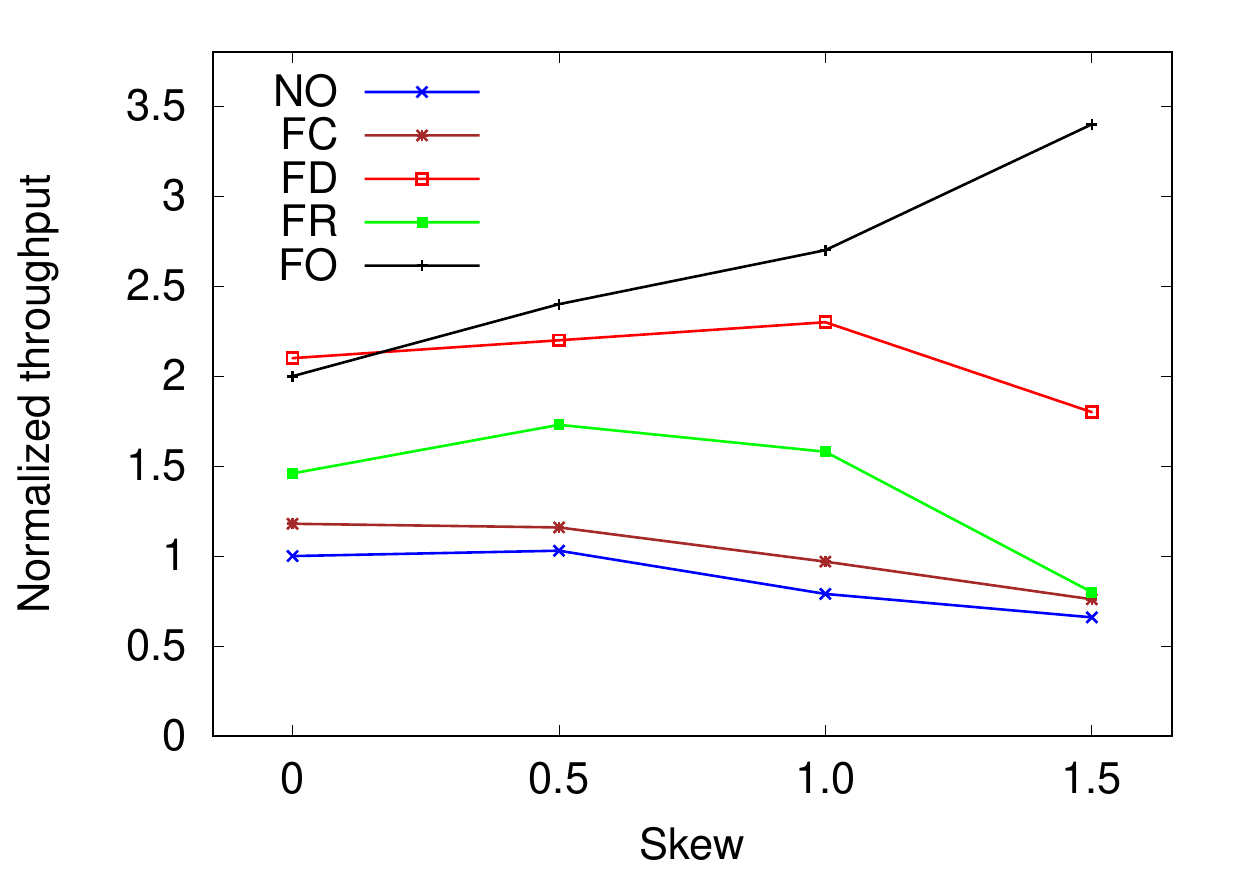}
		\caption{Data heavy workload}
		\label{fig:muppetDh}
	\end{center}
\end{subfigure}
\hfill
\begin{subfigure}{0.32\textwidth}
	\begin{center}
		\includegraphics[width=\textwidth,keepaspectratio=true]{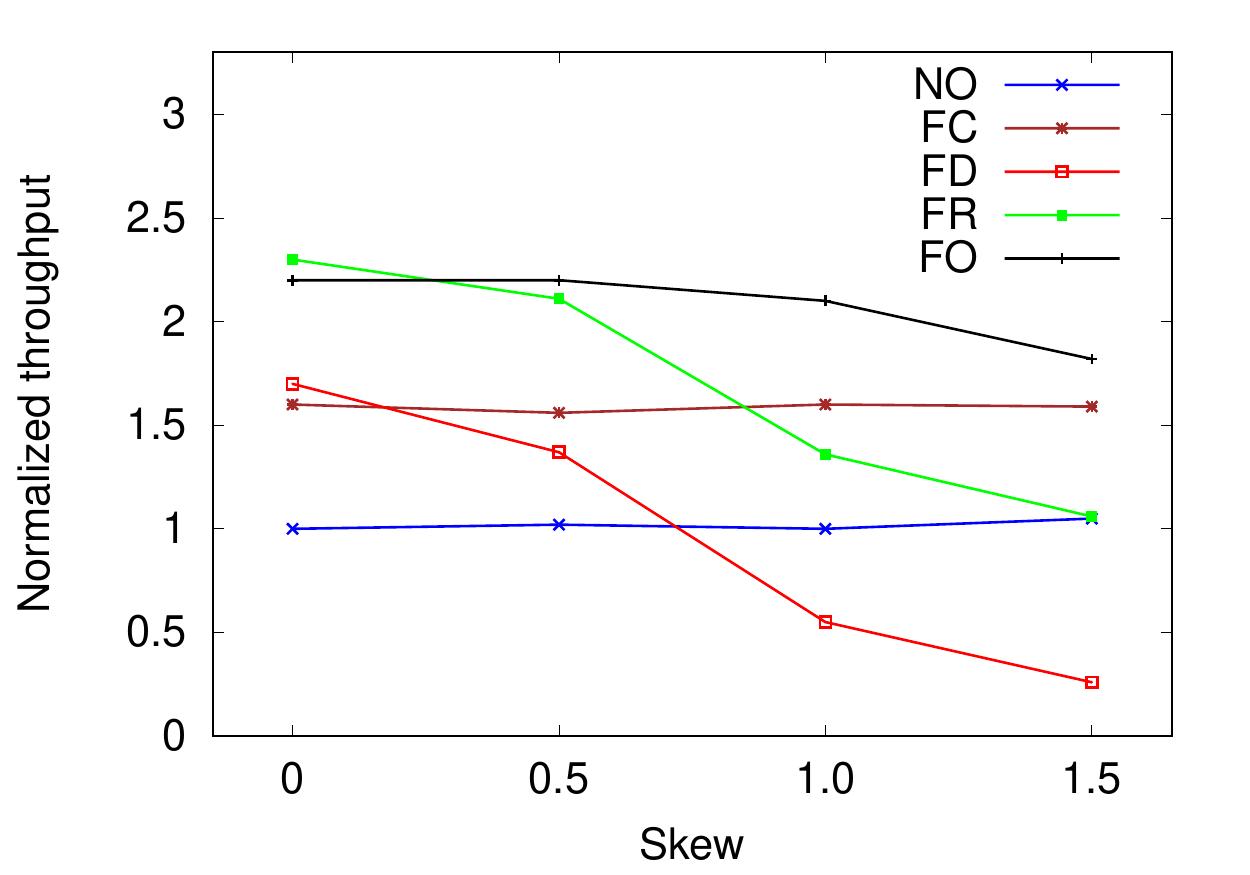}
		\caption{Compute heavy workload}
		\label{fig:muppetCh}
	\end{center}
\end{subfigure}
\hfill
\begin{subfigure}{0.32\textwidth}
	\begin{center}
		\includegraphics[width=\textwidth,keepaspectratio=true]{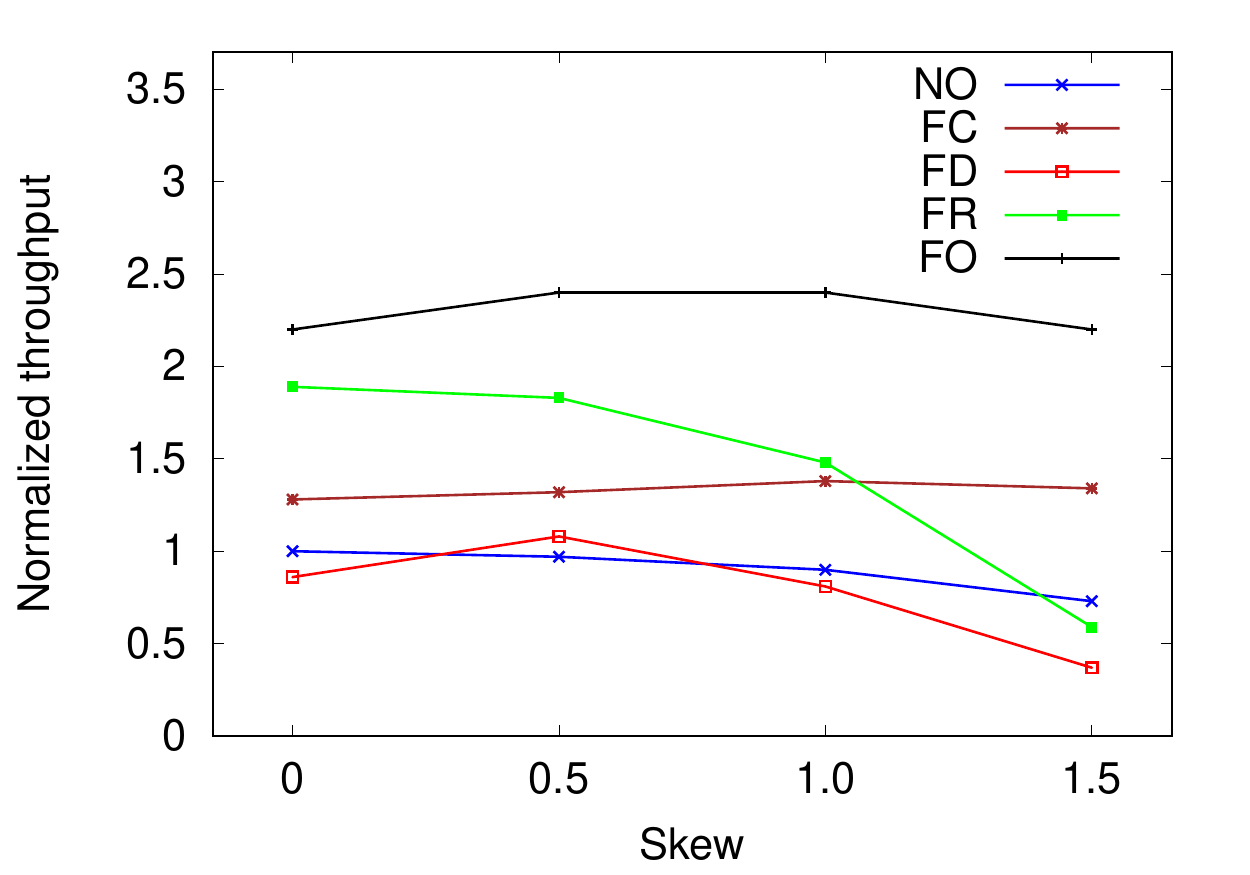}
		\caption{Data and compute heavy workload}
		\label{fig:muppetCo}
	\end{center}
\end{subfigure}
\caption{Muppet throughput comparison}
		\label{fig:muppet}
\end{figure*}

\subsection{Cache Implementation}
We use the Ehcache~\cite{ehcache} library for our cache implementation. Ehcache provides 
APIs for defining a composite cache that can be partially kept in memory and partially on 
disk. Users can also define the size of cache for the memory and disk components. We 
define our own eviction policy, for evicting cached items from memory, to disk cache by 
extending the \smalltt{AbractPolicy} class provided by Ehcahe. 

As mentioned earlier, data in the disk cache may actually be resident in memory as cached 
pages in the file-system buffer. Hence, reads from disk cache will incur file-system 
overhead, but may not incur actual disk access overhead, which can be very high for 
random I/O on hard disks.

\eat{
\begin{figure}
\begin{scriptsize}
\begin{verbatim}
preMap(docId,document) {
  spotContextList = {}
  for each spot in document.getSpots() {
      spotContext = 
                getContextRecord(spot, document)
      //prefetch submit
      prefetch(spotContext.key) 
      spotContextList.add(spotContext)
  }
  postMapQueue.put(spotContextList)
}
\end{verbatim}
\end{scriptsize}
\caption{Program rewritten using prefetch}
\label{fig:preRun}
\end{figure}
}

\subsection{Implementation of preMap}
\label{sec:prefetch}

As described in Section~\ref{sec:other_opt}, we have extended the Hadoop MapReduce, Spark 
and Muppet APIs to allow the user to write a \smalltt{preMap} function that can issue 
prefetch requests. 

To illustrate the use of the API we modify the entity annotation program in 
Figure~\ref{fig:mapEntity}. The program using prefetching is shown in 
Figure~\ref{fig:opt}. The \smalltt{submitComp} function is a prefetch call that 
internally adds a $(f,p,k)$ data item to a \smalltt{Prefetch} queue and returns. 
Process threads read data from the queue and an optimizer module decides whether to issue 
a data request or compute requests. Depending on the type of request the data item is 
added to corresponding data or compute queues. 
Once the function is computed (either at the data node or the compute node) the result 
is added to the \smalltt{resultHashMap} and can then be used later in map function.

The map function is then called on values from the queue and performs a blocking fetch 
using \smalltt{fetchComp}. 
If the value (data item or compute request) from the data node has already been 
fetched\eat{and is available},  it can be processed immediately; otherwise the compute 
node waits till the values become available.
The exact way in which the \smalltt{preMap} function is invoked depends on the
system on which it is implemented.

The invocation of \smalltt{preMap} function for different systems is 
implemented as follows

\begin{itemize}
 \setlength{\parsep}{0pt}

\item \textit{Hadoop}:
The Hadoop Mapper class uses a \smalltt{setup} function to initialize the Mapper and a 
\smalltt{run} function to read input and call the map function. 
We modify the \smalltt{setup} function in the Mapper class to start a 
new thread, from which the preMap function is called on each input data item. 
The user code implementing the \smalltt{preMap} function can 
issue prefetch requests from the data store.
Once the input data item is processed by preMap, it is added to a \smalltt{MapInput} queue
that we create. The \smalltt{run} function in the Mapper is modified to 
read values from the \smalltt{MapInput} queue (instead of 
directly from the input), and it then calls the \smalltt{map} function. 

\item \textit{Spark}:
 In Spark, \smalltt{map} or \smalltt{flatMap} functions may be used to operate on RDDs 
 similar to the map function in Hadoop. The user can create functions that need to be 
 called for flatMap or map. We extend the Spark RDD to include additional functions 
 \smalltt{flatMapWithPremap} and \smalltt{mapWithPremap}. 
 The parameters to {flatMapWithPremap} and \smalltt{mapWithPremap} are two user defined 
 functions (wrapped in a class). The user defined class implements the
 \smalltt{call(t, asyc)}  function, to define how the input tuple \smalltt{t} 
 will be processed. The \smalltt{async} object can be used to submit prefetch requests in 
 \smalltt{preMap} or retrieve results in \smalltt{mapWithPremap/flatMapWithPremap}.

\item \textit{Muppet}:
In the Muppet stream processing framework, the \smalltt{MapUpdat\-ePool} class reads 
values from the input 
stream and adds it to a Map queue for processing. We extend the framework by creating a 
new thread in the constructor of \smalltt{MapUpdatePool} for prefetching. The 
prefetching thread reads the input from the stream, adds it to a prefetch queue 
and adds the tuples to the Map queue in \smalltt{MapUpdatePool}. 
Another thread reads values from the prefetch queue and issues prefetches. 
\end{itemize}

Some preprocessing on the raw input data may be needed to decide what data items are to 
be prefetched. This processing would be repeated at the \smalltt{map} function also. In 
the entity annotation example shown in Figure~\ref{fig:opt}, the 
\smalltt{spots = document.getSpots} needs to be repeated. As an optimization, we 
implement another extension where the prefetch thread provides the preprocessed data to 
be consumed by a \smalltt{postMap} function instead. The user can provide implementations 
for \smalltt{preMap} and \smalltt{postMap}.

\subsection{Compute Request Implementation}
\label{app:compute_req_impl}

We have implemented libraries for the optimizations described in this paper in 
Java, with Apache HBase as the datastore. 
We use the endpoint coprocessor in HBase to perform computations at the data nodes. 

Note that invocations to the data store need to be done for batches of requests,
rather than for individual requests.   This can be done by creating a batched
form of the function, which takes a set of requests instead of a single request.
The decision on what to compute and what to return as data is made by the batched
version of the function.  Thus, no changes need to be made to the data store.

HBase stores tabular data partitioned into regions. Each data node may have one or more 
regions. Coprocessor calls to the HBase need to define a start key, an end key, the table 
name on which the coprocessor will be invoked and the coprocessor class. The HBase API 
sends the request parameters to each region that has keys in the range [startKey,endKey] 
and invokes the function calls on that region. The results obtained from each region  
are then accumulated by the caller.

Since HBase provides API to get the data node for each key, at each compute node 
we maintain one batch of requests for each data node. Once the batch for a data node fills up,
we send the batched request to the data node using coprocessor execution calls.

In our coprocessor calls, we need to pass the $(k_i,p_i)$ pairs along with the necessary 
statistics as request parameters. In case there are $r$ regions at one data 
node all the $(k_i,p_i)$ pairs will need to be sent $r$ times (once for each region) 
if the default HBase API is used. We instead provide a wrapper API which takes input 
$(k_i,p_i)$ pairs, the table name and the coprocessor class. Our API sends requests to 
regions such that only the $(k_i,p_i)$ pairs for which $k_i$ is in the range of a region 
are sent to that region.

\subsection{Bandwidth Estimation}
\label{app:bandwidth}

To measure the effective network bandwidth available for each compute node and data node, 
we send data packets from all the compute nodes to all the data nodes and vice versa and 
measure the time taken under load.  $netBw_i$ is the aggregate of the network 
bandwidth at node $i$. Bandwidth estimation is done once during the initial setup and the 
value of bandwidth is used for optimizations computations. 
The network bandwidth measurement step allows our framework to be used even if
the bandwidth to some nodes (e.g. intra-rack bandwidth) is different from the 
bandwidth to others (e.g. inter-rack bandwidth).  In this case, we compute
the average bandwidth across all destinations, reflecting the fact that communication 
will be distributed across all the destinations.

In the case of a dedicated network, the bandwidth will not change during a computation
except due to failures.
However, in a shared infrastructure, bandwidth may change.
To the best of our knowledge, there is no non-intrusive method to measure network bandwidth while the 
functions are running. Any attempt to measure bandwidth by sending additional packets 
will reduce the performance of the program, while measuring rate of arrival of requests/results
will not be able to separate the network bandwidth from processing bandwidth.
However, if desired the network bandwidth can be measured and updated at runtime at the
cost of affecting performance temporarily, and we can use the updated bandwidth in our 
computations. 

\section{Synthetic Workload Muppet}

For Muppet we compare the performance, in terms of the throughput (number of input tuples 
processed per unit time), for  NO, FC, FC, FR and FO. We consider the throughput to be 
one tuple per unit time for NO technique and normalize throughput for other techniques 
and skews. For the results of these experiments, higher numbers indicate better 
performance. 

The normalized throughput for data heavy workload is shown in Figure~\ref{fig:muppetDh}. 
The throughput of FD is comparable to that of FO at z=0 but the  throughput of FD 
decreases with skew because some data nodes process a lot of requests,  while the 
throughput of FO increases because of caching. Similarly, the performance of NO,FC and FR 
decrease with increase in skew. For this workload as well as the other workloads, FC 
performs better than NO because of prefetching/batching optimizations. 

The normalized throughput for compute heavy workload is shown in 
Figure~\ref{fig:muppetCh}. 
FR performs better than FO at less load, but throughput of FR reduces significantly with 
increase in skew because a few data nodes do most of the computations. FO provides the 
best performance at higher skew.
As with MapReduce the performance of FO decreases with increase in skew since most 
requests are for data items already in cache and hence function computation is done 
locally; the data nodes remain underutilized.

The normalized throughput for the compute and data heavy combined workload is shown in 
Figure~\ref{fig:muppetCo}. For this workload, FO performs better than other workload at 
all skews since it can balance the CPU and network loads and cases frequent items.

}
\end{document}